\documentclass[fleqn,usenatbib]{mnras}

\usepackage{newtxtext,newtxmath}

\usepackage[T1]{fontenc}

\DeclareRobustCommand{\VAN}[3]{#2}
\let\VANthebibliography\thebibliography
\def\thebibliography{\DeclareRobustCommand{\VAN}[3]{##3}\VANthebibliography}

\usepackage{graphicx}	
\usepackage{xspace}

\newcommand{\pc}{\,\mathrm{pc}}
\newcommand{\Msun}{\,\mathrm{M}_{\odot}}

\newcommand{\Myr}{\,\mathrm{Myr}}
\newcommand{\Myrs}{\xspace\mathrm{Myrs}}
\newcommand{\yr}{\,\mathrm{yr}}
\newcommand{\K}{\,\mathrm{K}}
\newcommand{\yrs}{\xspace\mathrm{yrs}}
\newcommand{\kms}{\,\mathrm{km\,s}^{-1}}
\newcommand{\gocm}{\,\mathrm{g\,cm}^{-3}}
\newcommand{\ev}{\,\mathrm{eV}}

\newcommand{\INL}{{\sc RT-Infall24}\xspace}
\newcommand{\INH}{{\sc RT-Infall23}\xspace}
\newcommand{\INVH}{{\sc RT-Infall22}\xspace}

\newcommand{\NoINL}{{\sc NoRT-Infall24}\xspace}
\newcommand{\NoINH}{{\sc NoRT-Infall23}\xspace}

\newcommand{\NONRT}{{\sc NoRT}\xspace}
\newcommand{\RT}{{\sc RT}\xspace}

\newcommand{\ramses}{{\sc Ramses}\xspace}
\newcommand{\ramsesrt}{{\sc Ramses-rt}\xspace}

\newcommand{\tinf}{t_{\rm I}}
\newcommand{\tAGB}{t_{\rm AGB}}

\defcitealias{calura19}{C19}

\newcommand{\secref}[1]{Section~\ref{#1}}
\newcommand{\figref}[1]{Figure~\ref{#1}}
\newcommand{\tabref}[1]{Table~\ref{#1}}
\newcommand{\equref}[1]{Equation~\eqref{#1}}

\title[Ionising feedback effects on star formation ]{Ionising Feedback Effects on Star Formation in Globular Clusters with Multiple Stellar Populations}

\author[A. Yaghoobi et al.]{
A. Yaghoobi,$^{1,2,3}$\thanks{E-mail:a.yaghoobi@iasbs.ac.ir}
J. Rosdahl$^{2}$,
F. Calura$^{4}$,  
P. Khalaj$^{1}$,
H. Haghi$^{1}$
\\
$^{1}$ Department of Physics, Institute for Advanced Studies in Basic Sciences (IASBS), 444 Prof. Yousef Sobouti Blvd., 45137-66731, Zanjan, Iran \\
$^{2}$ Univ Lyon, Univ Lyon1, Ens de Lyon, CNRS, Centre de Recherche Astrophysique de Lyon UMR5574, F-69230, Saint-Genis-Laval, France\\
$^{3}$ Department of Physics, Faculty of Sciences, Ferdowsi University of Mashhad, Mashhad, 91775-1436, Iran \\
$^{4}$ INAF - OAS, Osservatorio di Astrofisica e Scienza dello Spazio di Bologna, via Gobetti 93/3, I-40129 Bologna, Italy
}
\date{Accepted XXX. Received YYY; in original form ZZZ}

\pubyear{2021}

\begin{document}
\label{firstpage}
\pagerange{\pageref{firstpage}--\pageref{lastpage}}
\maketitle
%
 \begin{abstract}
Using 3D radiation-hydrodynamical simulations, we study the effects of ionising radiation on the formation of second-generation (SG) stars in Globular Clusters (GCs) with multiple stellar populations. In particular, we focus on massive ($10^7 \Msun$) and young (40-Myr old) GCs. We consider stellar winds from asymptotic giant branch (AGB) stars, ram pressure, gas accretion onto the cluster, and photoionisation feedback of binary stars. We find that the stellar luminosity is strong enough to warm and ionise the intracluster medium, but it does not lead to a significant gas expulsion. The cluster can thus retain the ejecta of AGB stars and the accreted pristine gas. In addition, efficient cooling occurs in the central region of the cluster within $50\Myr$ from the formation of first generation stars, leading to the formation of SG stars. Our results indicate that the inclusion of photoionisation does not suppress SG formation, but rather delays it by about $\sim10\Myr$. The time delay depends on the density of the pristine gas, so that a denser medium exhibits a shorter delay in star formation. Moreover, photoionisation leads to a modest decrease in the total SG mass, compared to a model without it.
 \end{abstract}


\section{Introduction}
Photometric and spectroscopic studies of globular clusters (GCs) over the past decade have led to the discovery that all old GCs, with a few exceptions such as Ruprecht 106 (e.g. \citealt{Villanova2013, Frelijj2021}), are composed of at least two stellar populations with different chemical compositions. The stellar population having the same compositions as field stars at the same metallicity [Fe/H], is referred to as the first generation (FG) or first population \citep{bastian2018}. 

The second generation (SG) or second population, however, shows anomalies in the abundance of light elements, i.e. stars are enriched in He, Na, N and Al and are depleted in C and O \citep{minniti1993,carretta2009,gratton2013}. Such anomalous abundances in stars are found to be unique to GCs and an intrinsic property of them. In particular, it is found that almost all old clusters with a present-day mass larger than $\sim4\times10^4\Msun$ show an anti-correlation between the abundance of Na and that of O \citep{carretta2010}. The Na-O anti-correlation and the existence of multiple stellar populations (MSPs), can be considered as the definition of a bona fide GC (e.g. \citealt{carretta2010}). It can effectively separate GCs from their younger and less massive counterparts, namely open clusters, which do not show any sign of the Na-O anti-correlation \citep{Pancino2010, carretta2010, Bragaglia2013, Bragaglia2014}, and therefore are comprised of only a single population of stars. 

\par In a number of GCs, where resolution allows, the abundance anomalies are accompanied by splits in the colour-magnitude diagram (CMD). All these pieces of evidence signify self-enrichment in GCs, where SG stars have formed in a medium composed of pristine (unprocessed) gas\footnote{The gas that has the same chemical compositions as the material from which the FG stars are formed.}, mixed with products of CNO cycling and p-capture processes at high temperatures. The pristine gas can be left over from the formation of FG stars or accreted later on. In either case, it cannot have been polluted with the ejecta of supernovae (SNe) since most present-day GCs show very little spread in heavy elements such as iron \citep{renzini2015}. Stars in some GCs such as $\omega$~Cen exhibit significant enrichment by SNe. These cases are in minority and are believed to have an exotic origin, i.e. they could be remnants of satellite dwarf galaxies captured by the Milky Way or merging events (e.g. \citealt{Bekki2003}). 

\par The formation of the second generation population is still shrouded in mystery. Some scenarios suggest that they are multiple generations of star formation formed from different materials within star clusters \citep{BekkiKroupa2017, bekki2019}. FG stars whose winds are slow enough to be retained in GCs, and have the right chemical compositions to enrich the intracluster medium\footnote{This should not be confused with the intracluster medium in the context of galaxy clusters.} (ICM) and explain the present-day observed abundance patterns could be the origin of SG stars. Candidates for such FG stars are massive binaries ($m/\Msun\geq20$; \citealt{Mink2009,Szesi2018}), fast-rotating massive stars ($20\leq m/\Msun\leq120$; \citealt{decressin2007,DecCharbMey2007,krause2013}), supermassive stars ($m/\Msun\geq10^4$; \citealt{Denissenkov2014,Denissenkov2015}); asymptotic giant branch (AGB) stars ($4\leq m/\Msun\leq 8$; \citealt{D'Ercole2008,Conroy2011,D'Ercole2016,Bekki2017}).

\par Despite being successful in some aspects, all proposed scenarios suffer from several shortcomings. First, none of the scenarios are able to precisely reproduce the observed abundance patterns in GCs with MSPs \citep{bastian2015,renzini2015}. Second, present-day GCs possess approximately equal numbers of FG and SG stars. Assuming a canonical \cite{kroupa2001} IMF for stars, the ejecta of FG stars and available pristine gas are not sufficient to produce a large number of SG stars. This is referred to as the mass-budget problem. Unless one assumes very extreme conditions such as a binary fraction close to unity, or a significant loss of (only) FG stars ($90\%$ or more), this problem cannot be readily solved. For example, in the scenarios based on (fast-rotating) massive stars, a substantial amount of gas needs to be present in the GCs and the gas has to be expelled over an extremely short time scale of $\leq10^5\yr$. As a result, stars within the cluster cannot adjust their overall velocity distribution in response to the instantaneous or impulsive mass loss and the cluster can lose $\sim 90\%$ of its FG stars \citep{Khalaj2015, Khalaj2016}. Such short gas expulsion time scales can only be achieved by powerful agents such as hypernovae and dark remnants \citep{krause2013, Khalaj2016}. Third, one needs to fine-tune the timing of the formation of SG stars. For example in scenarios based on massive and very massive stars, the SG stars are assumed to be formed within a short time scale of about $<8.8\Myr$ \citep{krause2013} after the formation of FG stars, i.e. before they start exploding as type II SNe. In contrast, in the scenarios based on intermediate-mass stars (such as the AGB scenario) the second population cannot form until after $30\Myr$. The last point is also an issue for the scenario of the early-disc accretion set forth by \cite{bastian2013}. Although this scenario does not suffer from the mass-budget problem, the circumstellar discs around accreting stars need to survive for up to $\sim10\Myr$ \citep{bastian2013}. In harsh environments, such as GCs with a dense core and strong feedback \citep{calura2015}, it is not clear whether these discs can survive over such a timescale. Moreover, this scenario predicts a continuous spread of stars in the CMD, whereas the observed GCs show (two or more) distinct evolutionary tracks in the CMD, which is indicative of different episodes of star formation. Therefore, the formation of MSPs in GCs has not been completely solved and remains as an open question. 

\par Among proposed scenarios, the AGB scenario \citep{D'Ercole2008,D'Ercole2016} has received much attention in the literature and is the focus of the present paper. According to the standard model of the AGB scenario laid out in \citet{D'Ercole2016}, SG stars cannot form until all Type II SNe of the FG stars have gone off and the cluster has expelled the SN ejecta containing Fe-rich material. This corresponds to $t\approx40\Myr$ after the formation of the star cluster (FG stars), i.e. when the massive AGB stars begin injecting their enriched ejecta into the ICM. Concurrently, the cluster accretes pristine gas from the surrounding interstellar medium (ISM) as it moves through its host galaxy. To accrete the required amount of pristine gas for the SG formation, the gravitational potential of the cluster has to overcome the ram pressure exerted by the ISM on the ICM \citep{lin2007,Conroy2011}. \citet{lin2007} showed that if a cluster has a central velocity dispersion greater than both the sound speed of the ISM and the relative cluster-ISM speed, gravity dominates the ram pressure and the cluster can accrete a significant amount of the ISM. The amount of pristine gas accumulated in such a cluster depends on the properties of the ISM and the cluster \citep{naiman2011}.

\par Recently, \citet{calura19}, hereafter \citepalias{calura19}, conducted a comprehensive study of the formation of SG stars in a very massive ($10^7\Msun$) cluster moving in the ISM based on the standard AGB scenario \citep{D'Ercole2016}. In addition to the ram pressure addressed by previous studies, they took into account other physical processes such as radiative cooling, self-gravity, star formation and stellar winds from AGB stars. In particular, they utilized three-dimensional (3D) hydrodynamic simulations, and demonstrated that a very massive cluster can both retain its AGB ejecta and accrete pristine gas, forming a large number of SG stars. 
Intriguingly, in their simulations, two sub-populations of SG stars with different helium abundances were formed, which is in agreement with observations. In \citet{Yaghoobi2022} we extended the work of  \citetalias{calura19} to a range of cluster masses in and showed that a $10^6\Msun$ cluster could also form a massive SG. However, this is not the case for a $10^5\Msun$ cluster, due to its weaker gravitational potential. After performing approximate corrections for long-term dynamical evolution, we found a positive correlation between the SG fraction and the cluster mass in the simulations done in the high-density ($10^{-23}\gocm$) medium. Moreover, we found a similar correlation for the maximum He enhancement of SG stars. Both these results are in agreement with observations (e.g. \citealt{Lagioia2019,milone2020}). Using a similar setup to ours, \citet{lacchin21} investigated the role of Type Ia SN explosions in the SG formation and found that these explosions are not able to significantly limit star formation in this high-density medium. All of these findings, imply that the formation of SG stars in the AGB scenario \citep{D'Ercole2016}, needs to occur in dense environments. In spite of this milestone, there are still more physical processes to be incorporated in the simulations, one of the most important ones being related to star formation and the associated stellar radiation. 

\par Massive stars in star clusters radiate with high luminosities at young cluster ages. The emitted  photons get absorbed by the ICM gas. This heats up the ICM and increases its pressure \citep{Conroy2011,gavagnin2017,chantereau2020}. As a result, radiation may play a non-negligible role in the gas expulsion from clusters and inhibiting the formation of SG stars. \citet{chantereau2020} showed that stellar radiation of present-day GCs can be enough to expel the ICM, including the gas accreted from the ISM and stellar winds, from the cluster. According to the outcome of their simulations, the radiative heating along with ram pressure can explain the observed negligible amount of gas in the core of present-day GCs. 
Moreover, \citet{Kroupa2018} propose that the ionising feedback of O stars can explain the formation of multiple coeval populations separated in age by about $1\Myr$ in very young clusters with a mass range of [$0.6$-$20$]$\times10^3\Msun$. According to this model, the stellar feedback is able to ionise and expel the residual gas from the centre of the cluster and reduce the FG formation. But the infall of molecular gas can resume star formation within the cluster and form another population.
For a very massive cluster, \citet{Wunsch2017} also show that self-shielding against ionising radiation can occur before $3.5\Myr$, resulting in an efficient cooling and then the formation of subsequent generations of stars.

\par In the case of FG formation from low-mass clouds, \citet{gavagnin2017} showed that photo-ionisation feedback is able to expel most of the gas within $3\Myr$ and effectively lowers the star formation efficiency. In addition, it considerably reduces the stellar density of the cluster and limits gas accretion on very massive stars. As a result, star formation is suppressed a few $\Myrs$ after the cluster formation and the high-mass end of the stellar mass function is suppressed due to radiation. On the other hand, stellar spectral synthesis models \citep{BPASS2017,S99,Conroy2009} show that the luminosity of a population of stars decreases by several orders of magnitude within the first $\sim50\Myr$. This fact raises the question of whether star formation in the presence of ionising feedback can be triggered again after this time.
Using a simple model, \citet{Conroy2011} estimate the role of radiative heating in the formation of SG stars. They predict that only after a delay of several $100~\Myrs$ can the gas within a young GC cool down to ideal conditions which trigger SG formation. Such a delay, caused by radiative heating, could pose a severe problem for the AGB scenario \citep{bastian2018}.

However, the cooling process depends on gas properties that change over the cluster lifetime. For example, the gas density within the core of a massive cluster is expected to increase after $\approx 40\Myr$ \citepalias{calura19}, due to the winds of the AGB stars and gas accretion from the ISM. Therefore, an efficient cooling process might also occur before $100\Myr$. 

\par The present paper is a follow-up to \citetalias{calura19}. We investigate the role of stellar radiative feedback in the formation of SG stars in the context of the AGB scenario. We study the role of ionising radiation in the gas accretion and the retention of stellar winds in GCs. The paper is organized as follows. In \secref{sec:setup} we describe the numerical methods and main assumptions that we use for our simulations. In \secref{sec:results}, we present the results of our two simulations for different ISM densities and finally in \secref{sec:discussion} we discuss and summarise our results. 
\begin{table}
	\centering
	\begin{tabular}{lcccccc} 
		\hline
		Simulation & M$_{{\rm FG}}$ \ ${\rm [M_{\odot}]}$  & $\rho_{\rm pg}$ \ $[\rm g \ cm^{-3}]$  & $\tinf$ \ $[\rm Myr]$ & RT  \\		\hline
		\INL  & $10^7$ &  $10^{-24}$ &   21 & On\\
		\INH{} & $10^7$  & $10^{-23}$&  1 & On\\
		\NoINL{} & $10^7$ &  $10^{-24}$ &  21 &Off \\
		\NoINH & $10^7$ &  $10^{-23}$&  1& Off \\ 
		\hline
	\end{tabular}
 	\caption{Main simulation parameters. Column designations: M$_{\rm FG}$ is the mass of the cluster; $\rho_{\rm pg}$ is the ambient density;  $\tinf$ is the time at which the infall reaches the centre of the cluster (\secref{sec:init}) at the time reference of this paper ($39\Myr$ after the FG formation).}
	\label{tabl1}
\end{table}
%
\section{Model description}\label{sec:setup}
We perform our simulations using \ramsesrt \citep{rosdahl2013,Rosdahl2015}, a radiative hydrodynamics extension of \ramses \citep{teyssier2002}. The code uses a second-order Godunov scheme to solve the Euler equations and a particle-mesh solver to compute the dynamical evolution of particles.  The setup of our simulation is almost the same as \citetalias{calura19}, but with the difference that we include ionising radiation using the radiative transfer methods described in \citet{rosdahl2013}. To recap, we simulate a cluster with a mass of $10^7\Msun$ and a half-mass radius of $30\pc$ placed in a homogeneous gas with a uniform density. We consider two models with different ISM (gas) densities of $10^{-24}\gocm$, corresponding to a typical dwarf galaxy \citep{marcolini2003}, and a density of $10^{-23}\gocm$, corresponding to star-forming galaxies at high redshifts \citep{wardlow2017}. We refer to these models, including the ionising feedback, as \INH and \INL, respectively. For a better comparison between the results of radiative hydrodynamics simulations (\RT) and those of non-radiative (\NONRT) cases, we simulate two models without radiation, referred to as \NoINL and \NoINH. \tabref{tabl1} lists our main simulations and their associated parameters.

In \citetalias{calura19}, the simulations volume was a cube with a width of $162\pc$. 
In the current simulations, we find that this volume was too small for the results to converge, due to the photoionisation expansion of gas beyond the boundaries. Therefore, we consider a larger box of $256\pc$ so that the regions affected by radiation in our simulations will be better confined within our computational volume. The total volume of our simulation box is $256^3\pc^3$. We mesh this box into a coarse grid with a maximum cell size of $\Delta x_{\rm max}= 2\pc$, corresponding to the minimum level $l_{\rm min}=7$ in the \ramses code, and a minimum cell size $\Delta x_{\rm min}= 0.5\pc$, i.e. $l_{\rm max} = 9$. The code adaptively refines when the total mass inside a cell exceeds $50\Msun$. We find this resolution and refinement scheme to be sufficient for our results to convergence. For the sake of simplicity, we model the gravity from the FG stars as a static \citep{plummer1911} potential and omit the dynamical evolution of the FG stars. The total mass of FG stars therefore does not change over time during our runs. 
%
\subsection{Initial conditions}\label{sec:init}
Following the AGB model presented in \citet{D'Ercole2016}, we start our simulations at $\tAGB=39\Myr$ after the formation of the cluster. At this time, all SN explosions of FG stars have subsided, resulting in a hot bubble of diffuse gas formed around the cluster. Therefore the bubble is assumed to have already reached a stagnation radius from the cluster center, where the expansion velocity of the bubble is equal to the velocity dispersion of the ISM. This radius is denoted by $R_{\rm eq}$  and depends on the properties of the cluster and ISM, such as cluster mass, ISM density and velocity of the cluster with respect to the ISM (see \citetalias{calura19} for more details). Thus the cluster is not initially in direct contact with the ISM as the region inside the hot bubble ($R<R_{\rm eq}$) is filled with a low-density gas ($\rho \sim 10^{-31} \gocm$) formed by SNe ejecta and is in hydrostatic equilibrium with the cluster. By ignoring the dynamical instabilities of the bubble such as radiative Rayleigh–Taylor instability \citep{Jacquet2011,Yaghoobi2018MNRAS}, we can assume that the cluster traverses the radius of $R_{\rm eq}$ and arrives at the interface of the ISM at time $t_{\rm I}$, after which it starts accreting pristine gas from the ISM. In our setup, it is assumed that the relative speed of the cluster with respect to the ISM is $23\kms$ and remains constant. This is a typical value for clusters in strongly interacting systems such as the Antennae galaxies \citep{Whitmore1999AJ,Gilbert2007}. Following the approach explained in \citetalias{calura19}, we  estimate the values of $t_{\rm I}$ to be $\sim60\Myr$ and $\sim40\Myr$ for the \INL and \INH cases, respectively. These time scales are measured from the formation time of the cluster (FG stars). Given that all simulations start at $39\Myrs$ after the FG formation, the cluster reaches the ISM at simulation times of $21\Myr$ (low density) and $1\Myr$ (high density).

We perform the simulations in the reference frame of the cluster, i.e. the centre of the cluster always lies still at the centre of the box and ISM gas moves in from the side at the assumed velocity of the cluster after the time of $t_{\rm I}$.
%
\subsection{Stellar winds of FG stars}
In general, the properties of stellar winds are complex functions of the mass, metallicity, and age of the stars they originate from, i.e. the AGB stars in our case. Our approach to factor the wind ejecta of the AGB stars into our simulations is as follows. Starting from $\tAGB=39\Myr$ after the formation of FG stars, we progressively inject enriched material into the simulation cells, at a rate which depends on the age and mass of the FG stars in each cell. The rate is given by the following relation
\begin{equation}\label{AGB_ej}
\dot{\rho}_{ \rm AGB}=\alpha \rho_{*,\rm FG},
\end{equation}
where $\rho_{*,\rm FG}$ is the FG mass density as a function of radius from the cluster centre. and $\alpha$ is the specific injection rate. For a typical cluster with a canonical \citet{kroupa2001} IMF, it is given by $\alpha(t)=0.065t^{-1.01}\yr^{-1}$ \citep{ciotti1991}, where $t$ is the time after the FG formation measured in $\yrs$. Following \citetalias{calura19}, the rate of energy injection into the cells is assumed to be:
\begin{equation}\label{EAGB}
S=0.5\alpha \rho_{*,\rm FG}\left(3\sigma^2+v^2+v_{\rm wind}^2\right),
\end{equation}
where $\sigma$ is the one-dimensional velocity dispersion of the cluster, $v_{\rm wind}$ is the wind velocity of the AGB stars and $v$ is the local gas velocity \citep{D'Ercole2008}. We assume a wind velocity $v_{\rm wind}=20\kms$ and a metallicity of $Z=0.001$. 
 
The ejecta from the AGB stars contain a significant amount of helium as well as products of CNO processes \citep{renzini2015,bastian2018}. However, we only focus on their helium mass fraction $Y$, varying from $Y(t=39\Myr)=0.36$ to $Y(t=102\Myr)= 0.32$ \citep{ventura2011}. We use a passive scalar to follow the He mass fraction of the gas in each cell. The mixing of the AGB ejecta and the accreted pristine gas (with $Y=0.246$) results in a gas with intermediate He abundances from which the SG stars can form.
%
\subsection{Star formation}\label{SF}
We use the star formation technique developed by \citet{Rasera2006}. 
Stellar particles are formed out of the gas using a standard \citet{Schmidt1959} law based on some specified thresholds. In our setup, when the gas has a temperature less than $2\times10^4\K$ and it has a converging net flow $(\nabla\cdot \textbf{\textit{v}}<0)$ stars are allowed to form.  To ensure that the gas in a cell has the required conditions to trigger the star formation, we also require that the local Jeans' length is smaller than 4 (finest) cell width, i.e. less than 2 pc. We set a minimum particle mass of ${\mathrm{m}_{*} = 0.1\Msun}$ and do not allow more than $90\%$ of the cell mass to be converted into a star particle at a given time. According to the resolution assumed in this study, this implies an effective density threshold of $\rho>6\times10^{-23}\gocm$ for star formation. In star-forming cells, gas is stochastically turned into star particles and they are placed at the centre of the cell. The mass of each stellar particle is determined as an integer multiple of the minimum particle mass, i.e. ${m_{\rm p}=N\mathrm{m}_{*}}$. Where $N$ is drawn from  a Poisson distribution, as described in \citet{Rasera2006},  so that on average particles form according to the \citet{Schmidt1959} law  with a star formation timescale of $0.1\ $Gyr. The velocity, metallicity, and He abundance of the star particles are also set to be the same as those of its natal gas. As in \citetalias{calura19}, the star formation timescale ($t_*$) is assumed to be $0.1 {\rm\, Gyr}$ in our setup.

The IMF of SGs is not known yet. However, it is commonly assumed that the mass of SG stars is lower than $8\Msun$ (e.g. \citealt{renzini2015, Khalaj2015, Khalaj2016,calura19,Yaghoobi2022}). In support of this assumption, it is found that the gravitational potential of FG stars can suppress the formation of massive stars and can truncate the IMF of SG stars \citep{bekki2019}. Therefore, we assume that the IMF of SG stars is bottom-heavy in our simulations.  As a result, SG stars do not turn into Type II SNe. Assuming otherwise yields some issues for the SG formation scenarios to explain the observed results, as strong SNe feedback from massive SG stars suppresses star formation \citep{calura2015} after a few $\Myrs$. This effect further reduce the number ratio of SG to FG stars, making the mass-budget problem more stringent. In addition, the gas available for star formation can be contaminated by  the SN ejecta of SGs, resulting in spreads in heavy elements which is in contradiction with observations.
%
\subsection{Radiative transfer}\label{sec:RT}
\begin{figure}
\centering
\includegraphics[width=\linewidth]{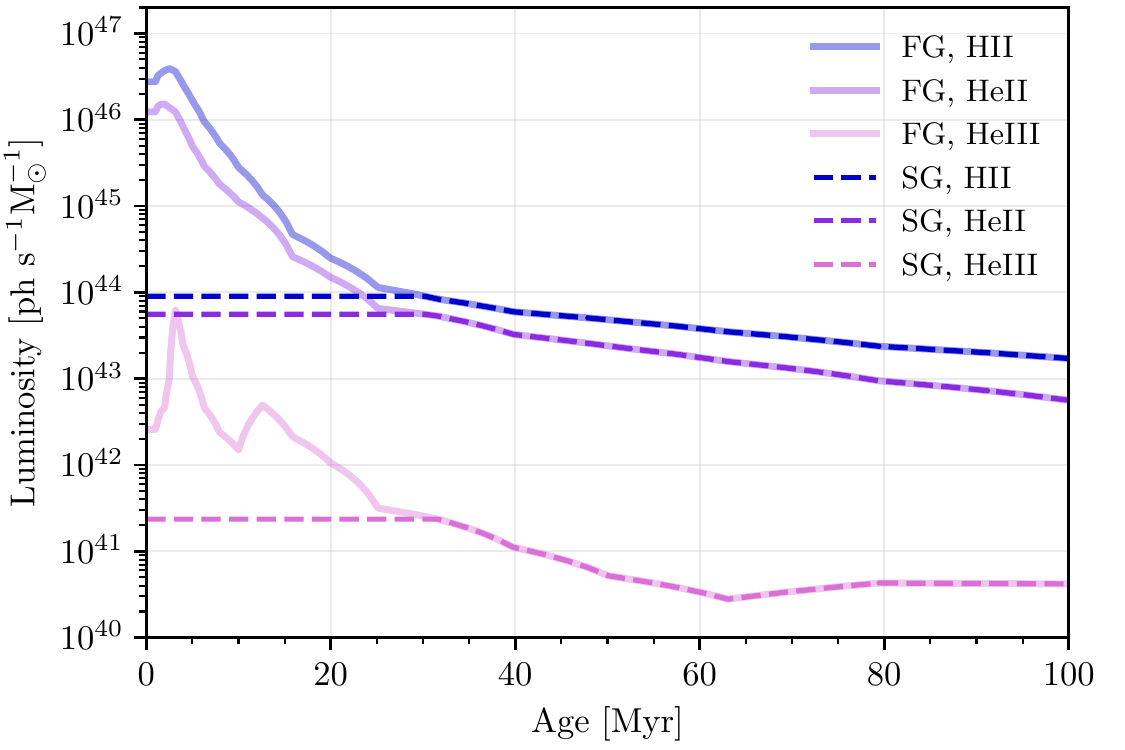}
\caption{The assumed SEDs for FG and SG stars. Based on the BPASS model, this figure shows ionising luminosities per solar mass in the three radiation groups (HI, HeI and HeII ionising photons) as a function of stellar age for first- (solid lines) and second- (dashed lines) generation stars at the metallicity 0.001. Note that we assume no stars with mass $<8\Msun$ for the second generation, hence the luminosities are constant in the first $\approx 30\ \Myrs$. }
\label{fig:SEDs}
\end{figure} 
\begin{table*}
	\centering
	\begin{tabular}{lcccccc} 
		\hline
		Photon group & $\overline{\epsilon}$\ ${\rm [\ev]}$  & $\sigma_{\rm HI}$ \ $[\rm cm^{-2}]$  & $\sigma_{\rm HeI}$ \ $[\rm cm^{-2}]$ & $\sigma_{\rm HeII}$ \ $[\rm cm^{-2}]$  \\		\hline
		UV$_{\rm HI}$   &  18.2  & $3.3\times10^{-18}$  & 0        &  0       \\
		UV$_{\rm HeI}$  &  33.0  & $6.4\times10^{-19}$  & $4.8\times10^{-18}$  &  0       \\
		UV$_{\rm HeII}$ &  61.3  & $9.2\times10^{-20}$  & $1.4\times10^{-18}$  &  $1.2\times10^{-18}$ \\
		\hline
	\end{tabular}
 	\caption{Properties of photon groups used in this study. Column designations: $\overline{\epsilon}$ is the mean photon energy; $\sigma_{\rm HI}$, $\sigma_{\rm HeI}$ and $\sigma_{\rm HeII}$ are cross-sections for ionisation of hydrogen and helium, respectively.}
	\label{tab:photonG}
\end{table*}
The \ramsesrt code follows the propagation of radiation and its interactions with the gas via photoionisation and heating. In this code, the radiative transfer equation is dimensionally reduced by taking the first two moments and using the M1 closure to describe the evolution of the radiation flux. The radiation interacts with  hydrogen (H) and helium (He) via non-equilibrium thermochemistry and  the non-equilibrium ionisation fractions of H and He are followed in every cell. The number of photons injected into every cell from stars is determined by spectral energy distribution (SED) models with respect to the mass, age, and metallicity of each stellar particles, while the photon properties (individual photon energy and cross-section) are set according to a blackbody radiation distribution.

In our simulations, ionising sources include both FG and SG stars. We model the SED of the FG and SG with the binary population and spectral synthesis (BPASS) code \citep{BPASS2017}.
Initially, the FG stars are assumed to have an age of $39\Myr$, a metallicity of $Z=0.001$ and a \citet{kroupa2001} IMF. SG stars are assumed to have the same metallicity and SED, except that their IMF is truncated at $m=8\Msun$, resulting in a constant luminosity of $9.0\times10^{43}\ {\rm photon}\ \Msun^{-1}\,{\rm s}^{-1}$ for the SG stars that are younger than $30\Myr$. Such a constant luminosity can also be obtained via other spectral models, such as the Starburst99 \citep{S99}, for a population with a bottom-heavy IMF.\figref{fig:SEDs} shows the assumed luminosities of FG and SG stars as a function of time in our simulations. We assume three photon groups (HI, HeI and HeII ionising photons) whose properties have been listed in Table \ref{tab:photonG}, according to the assumed SED. 

Owing to the explicit RT solver in \ramsesrt, the timestep length scales inversely with the speed of light, which at a full light-speed makes for orders of magnitude shorter (and therefore more) timesteps than in a pure hydro simulation. By adopting a reduced speed of light ($0.002c$) in our simulations, \ramsesrt provides us with a moment-based approach that reduces the computational cost. We find that our results have only a marginal dependence (if any at all) on the adopted value for the speed of light.

Besides photoionisation heating, our simulations also include radiative cooling and radiation pressure. We find that the radiation pressure has no effect on SG formation, i.e. all the effects of radiation are from photoionisation heating. The metal cooling process is based on the model described in \citet{sutherland1993cooling} for cooling of gas for H, He and metal lines in temperatures  $>10^4\ $ K.  For lower temperatures, metal cooling is assumed by  \citet{Rosen1995ApJ} with a temperature floor of $10^3\ $K. Finally, we consider a adiabatic index of $\gamma = 5/3$ for the ratio between internal energy and gas pressure. 
\begin{figure*}
\centering
\includegraphics[width=\linewidth]{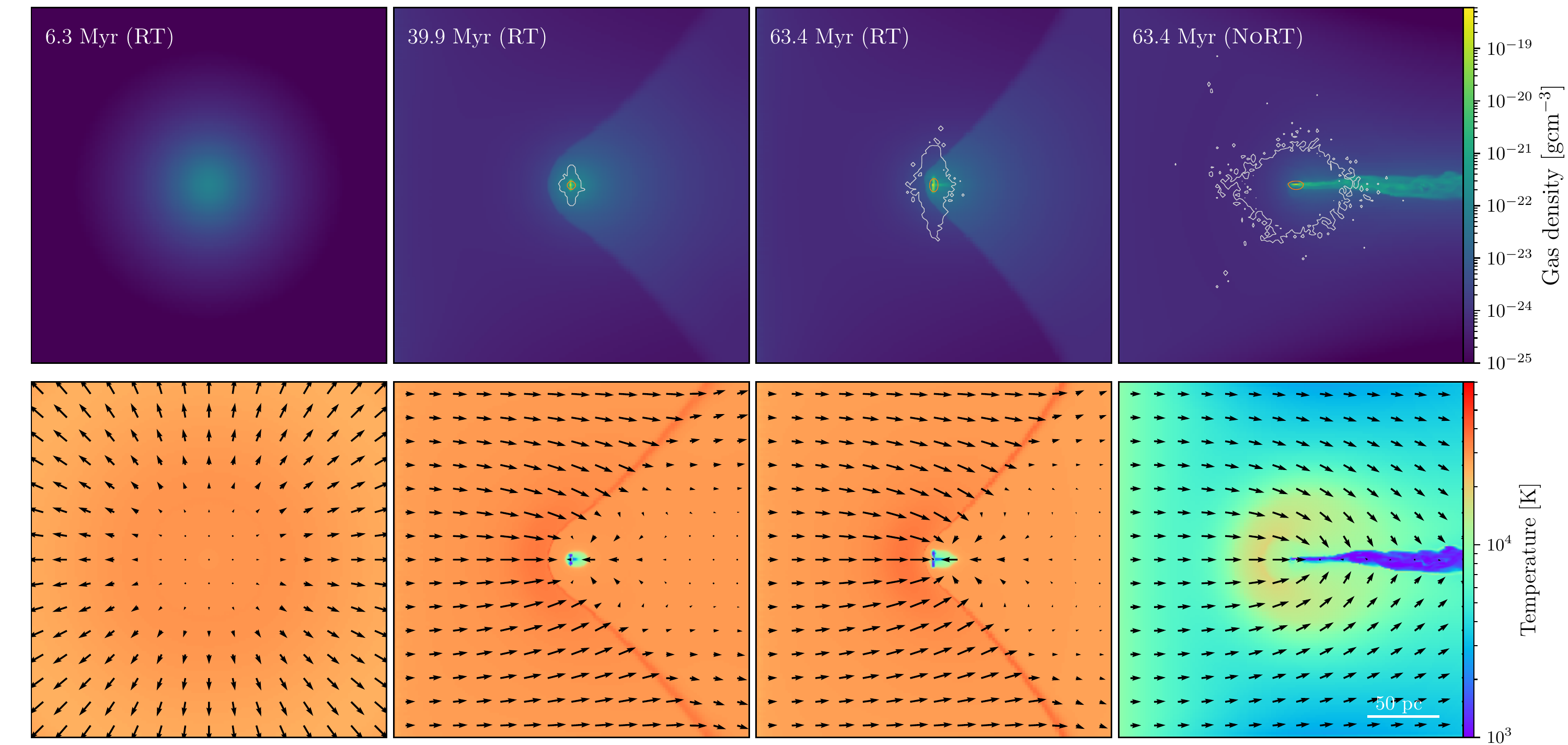}
\includegraphics[width=\linewidth]{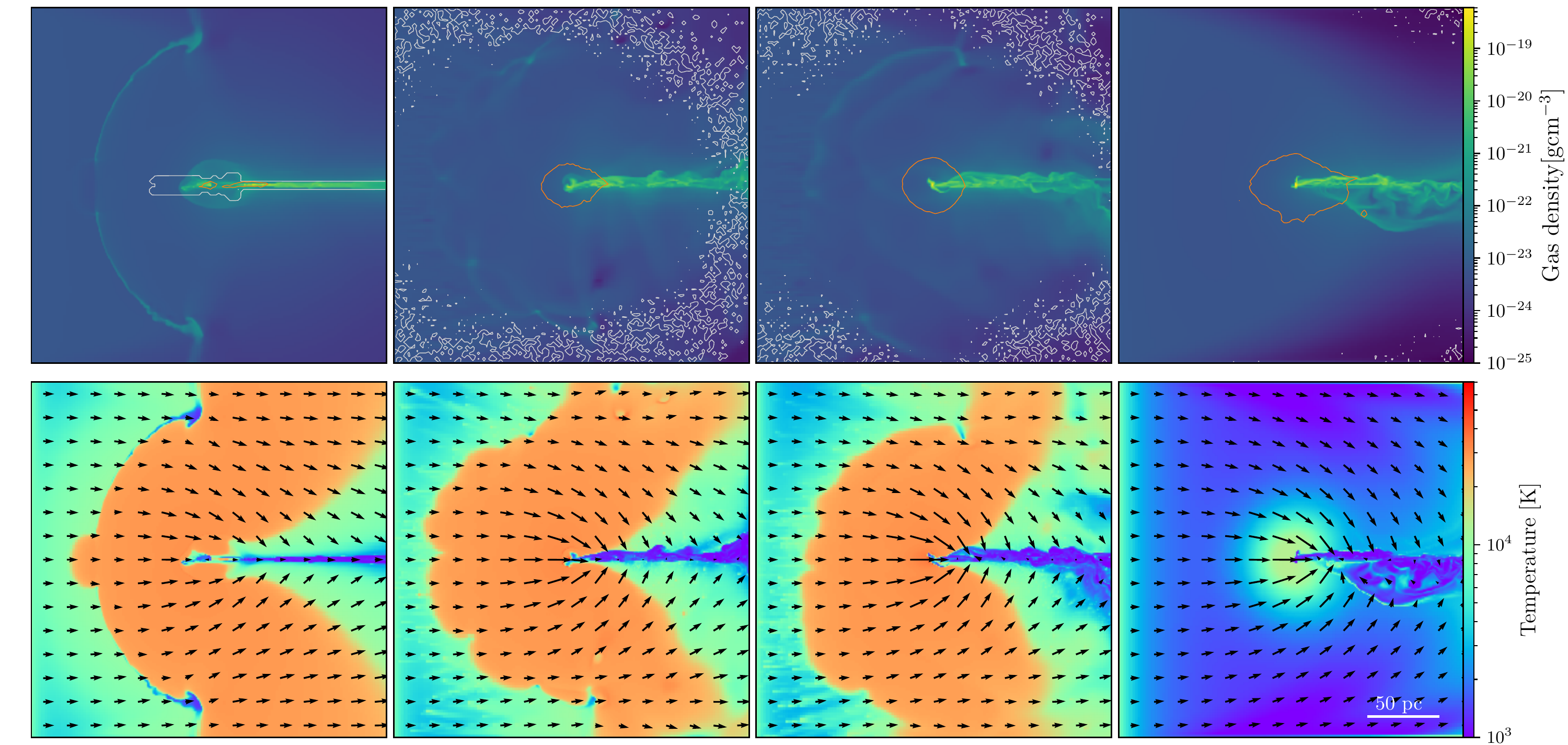}
\caption{Gas density (first and third rows) and temperature (second and fourth rows) maps for \INL (the top two rows) and  \INH (the bottom two rows). The figures show slices of the simulations in a plane intersecting the box centre in different snapshots and in the presence of radiation feedback. The last two columns compare the cluster at $t=63.4\Myr$ for the \RT and \NONRT cases. 
The white and orange contours show regions in which the SG stellar density are $10^{-6}$ (i.e. the extent of all SG stars) and $10^{-3}$ times the maximum central density (related to the \INH simulation), respectively.
The black arrows represent the gas velocity field. In the case of \INL, photo-ionisation heating pushes the gas out to large radii, preventing the formation of an accretion column beyond the cluster. The radiation feedback from the FG stars also delays the formation of the SG stars.}
\label{fig:IN}
\end{figure*}
%
\section{Results}\label{sec:results}
In this section, we present the results of our \RT simulations including the radiative stellar feedback of both generations of stars and compare them with their corresponding \NONRT cases. We investigate the role of radiation on the retention of He-rich stellar winds, gas accretion from the ISM, and SG formation.

Our simulations consist of two consecutive phases. In phase I, before the infall of pristine gas, the only injection of gas into the box is from AGB ejecta. This gas is immediately exposed to the ionising radiation from FG stars and as a result, the temperature and pressure increase everywhere inside the box.
Moreover, the FG radiation is able to expand the ICM via photoionisation heating. In phase II, the cluster reaches the ISM and moves through it.
In this phase, radiation in unison with ram pressure opposes the gravitational pull of the cluster (via photoionisation heating). The properties of the cluster and the ISM, such as mass and density, ultimately determine which effect dominates \citep{chantereau2020} and how much gas will be accreted. As stated earlier, our simulations start at the onset of the AGB ejecta, i.e. $\tAGB=39\Myr$ after the formation of the FG stars. They finish at $t=102\Myr$, at the start of the FG Type I SNe. For simplicity, all time instances stated in the following sections are measured from $\tAGB$ (and this have a final time of $64.3\Myr$).
\begin{figure*}
\centering
\includegraphics[width=\linewidth]{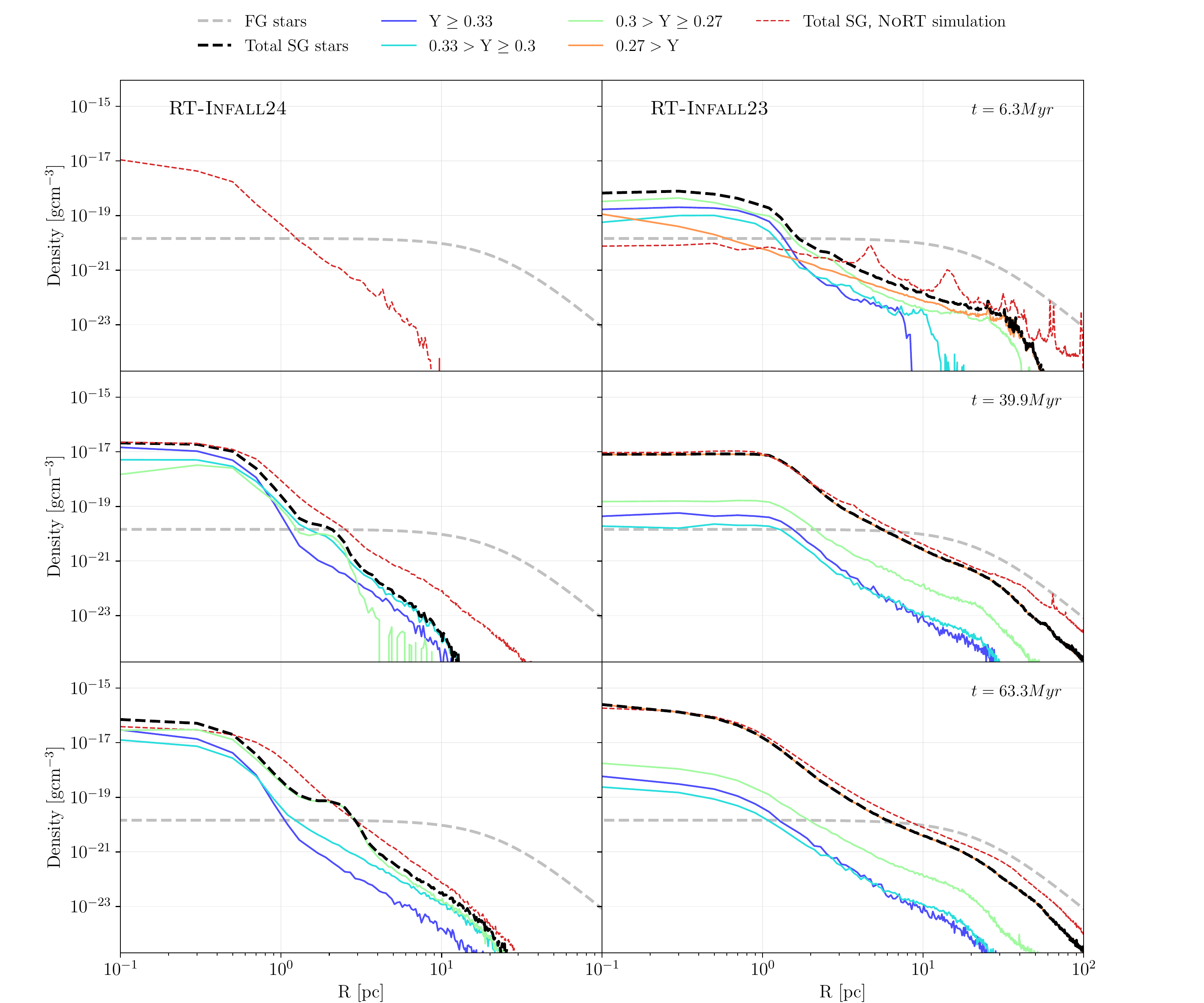}
\caption{Stellar density profiles of FG and SG stars at different times. For the FG stars only the overall profile is shown, which is the Plummer density profile. For the SG stars we show the total density profile but also split it into components with different helium abundances, as indicated in the legend. Moreover, the red dashed lines show the total SG profiles for the NoRT cases.}
\label{fig:profs}
\end{figure*}
%
\subsection{\INL}
\figref{fig:IN} illustrates the time evolution of the gas (and stellar) density, temperature and velocity fields of the \RT model over a time interval of $6.3-63.4\ \rm(according\ to\ the\ legend)\Myr$. The top two rows correspond to \INL and bottom rows correspond to \INH. For comparison, the last column on the right shows each simulation performed assuming the \NONRT model.  

\par The pristine gas reaches the center of the cluster at $t=21\Myr$ in \INL. The first snapshot (the top two panels of the first column) shows a density enhancement due to the injection of the He-rich stellar ejecta in the central region. Moreover, radiative heating has raised the gas temperature to over $10^4\K$ everywhere in the simulation box, leading to the gas expanding from the centre. Within the first $6\Myr$, cooling is not sufficiently strong to overcome photoionisation heating, hence no stars are formed. As the simulation progresses, the luminosity of the FG stars gradually drops (\figref{fig:SEDs}) while the gas injected into the box increases the central density considerably, which reduces the efficiency of the radiation in heating the gas. Combined with the fact that the cooling process depends directly on the density squared, this ultimately leads to more efficient cooling at the centre. As a result, the gas in the cluster centre cools down and SG stars start to form from He-rich gas at about $11\Myr$ in the \RT case. By comparison, formation of the SG stars begins at $2\Myr$ for the \NONRT case. Therefore, a delay in the star formation is the first effect of the ionising radiation on  SG formation. At this point, the gas is everywhere ionised except in the centre of the box. The formation of the SG stars leads to the emission of new photons. However we find that their role in heating the gas is negligible compared to the FG radiation, since the SG mass is much lower than the FG one (and because the SG is not assumed to contain massive stars, leading to low luminosities for the SG stars). Throughout the infall of pristine gas, the FG radiation has the effect of slowing He-poor gas out which slows down the infall gas velocity. This effect counteracts ram pressure in removing the gas reservoir from the cluster. However, the gas expansion due to photo-ionisation heating causes the cross section (effective surface area) of the gas to increase, making the ram pressure more effective in stripping the ICM. As the ISM passes through the cluster center and gets accumulated, a bow shock is formed behind the cluster.

At $t=39.9\Myr$ a distribution of SG stars is visible around the center of the cluster. As the simulation progresses, the efficiency of radiation in heating the ICM reduces even further. Moreover, the ejecta of the AGB stars become even more diluted with the accreted pristine gas, resulting in the formation of SG stars with lower He abundances. At the final snapshot of our simulation ($t=63\Myr)$, the total mass of SG stars has grown to $6\times10^5\Msun$. This is approximately $67$ percent of the total SG mass in the \NONRT case. \figref{fig:IN} shows that SG stars in the \NONRT case are more extended than in the \RT case. Moreover, in the \RT case the ICM is still too hot to form stars, except for the very central dense region of the cluster. A gaseous tail is formed down-flow in the \NONRT simulation, but the radiative feedback prevents this from forming in the \RT case.

\figref{fig:profs} shows density profiles of the FG and the SG stars from the center of the cluster in the same snapshots as in \figref{fig:IN}. The SG profile is split into components with different helium abundances but we also show the total SG profile, for the \RT (black dashed line) and \NONRT (red dashed line) cases. The bottom-left panel shows a slight difference between final SG profiles in both cases with and without radiation. The central density for both cases is greater than $10^{-17}\gocm$, which is a typical value in the present-day GCs (e.g. \citealt{renzini2015}). Moreover, the SG stars in both the \RT and \NONRT simulations are more concentrated than the FG stars in the central region (within a radius of $\approx 3\pc$), while they are characterized with ${\rm Y>0.27}$. 

\begin{figure*}
\centering
\includegraphics[width=\linewidth]{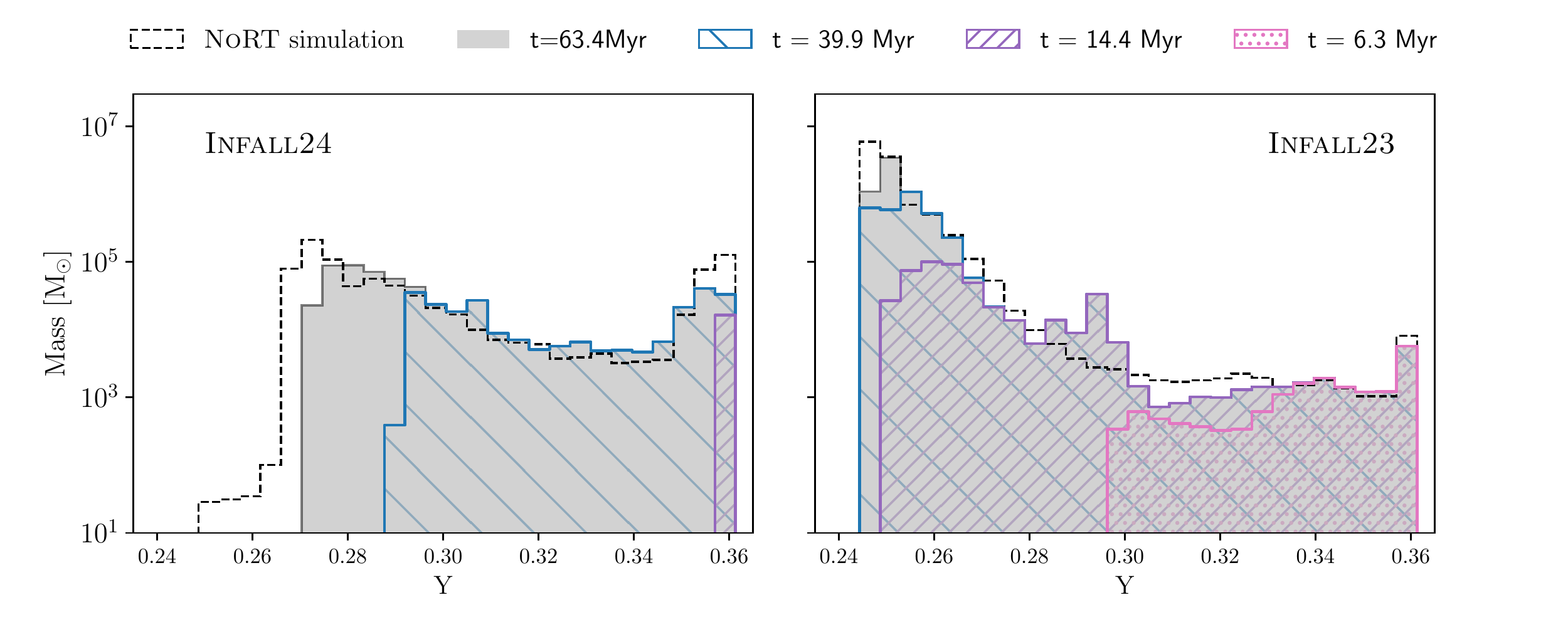}
\caption{Mass distribution of SG stars versus He abundance $Y$ at different times. The helium mass fraction of the SG stars varies between $Y=0.25$, which is the helium mass fraction of the pristine ISM gas (and FG stars), and $Y=0.36$, originating purely from the most massive AGB stars at the beginning of the simulations.}
\label{fig:distrib}
\end{figure*}

\figref{fig:distrib} illustrates the mass distribution of SG stars as a function their helium abundances at different times in our simulations. In the \INL simulation, as expected, a unimodal distribution takes shape  just before the infall reaches the center at $t=21\Myr$, reflecting that the SG stars are formed from pure AGB ejecta. As the simulation progresses and the AGB ejecta are mixed with pristine gas with $Y=0.246$, the distribution tends towards lower helium abundances.
So that another peak at the final time of $t=63.4\Myr$ is formed at $Y=0.27$, showing that younger SG stars are formed from very diluted AGB materials. Therefore two SG sub-populations with different He abundances can form in the low-density simulation.
%
\subsection{\INH}
We now describe the formation of SG stars for \INH, where the ISM density is ten times more than that of \INL. A denser medium yields a stronger ram pressure than in the lower-density model. However, the cluster can still overcome the ram pressure stripping and accrete a significant amount of pristine gas. In this model, the infall starts much sooner than in the \INL case ($1\Myr$ after the start of the AGB ejecta). This is due to the fact that in a denser ISM, a smaller bubble is formed by Type II SNe of the FG stars (equation 2 in \citetalias{calura19}). Therefore, the cluster reaches the pristine gas beyond this bubble in a shorter time.  

As depicted in the third row of \figref{fig:IN}, a compact population of SG stars has formed at the centre of the cluster after about $6\Myr$. This model forms a tail at $6.3\Myr$, though, it is not as well-established as the tail formed in the \NONRT case at the same time. Moreover, a relatively dense and symmetric shell is seen ahead of the cluster that is a result of the photoionisation-heated gas pushing against the incoming gas. This implies that the FG radiation is not strong enough to affect all the gas in the entire simulation box, in contrast to \INL. The bottom-left panel of \figref{fig:IN} shows that the gas temperature in the inner area of this shell is greater than $10^4\K$ and it is ionised while the gas in outer regions is much colder than to the pristine incoming gas. The motion of the pristine gas can be followed by the arrows on the temperature map in \figref{fig:IN}. It shows that as the incoming gas flows to the right, and accretes into a dense and cold tail behind the cluster at later times. Moreover, the white contours indicate that the stellar distribution of the SG stars in this model is more extended than in the \INL due to the denser infall. The broader distribution in this model can also be seen from the right column panels of \figref{fig:profs}. The first snapshot shows that almost all the stars have helium abundances higher than $0.27$, i.e. the gas fuelling the SG formation is enriched by AGB ejecta at this point. Interestingly, the SG stars in the \INH are more concentrated toward the centre than in the \NONRT simulation at the same time. The reason is that the stronger ram pressure in the latter mitigates the central collapse. The SG density is higher than the FG one at the core, but as the simulation progresses SG stars with lower helium abundance begin to dominate in the central region. A comparison between the final SG profile in the \NONRT and \RT simulations at the final snapshots reveals that the SG density in the presence of the ionising sources decreases slightly  in the outer regions. However, this difference in the central regions is none. As in the \INL simulation, the central density for both cases is still greater than $10^{-17}\gocm$, which is in agreement with observations (e.g. \citealt{renzini2015}). At the final snapshot, the SG stars dominate in the inner part of the cluster ($R\leq5\pc$), compared to the FG profile. This radius is  approximately twice larger than that where the SG stars dominate in the low-density simulation. 

From the right panel of \figref{fig:distrib} we can conclude that the final He distribution in our \INH simulation is insensitive to the inclusion of radiation feedback. As in the \INL simulation, two peaks appear at $Y = 0.36$ and $Y=0.25$ in both cases of the \RT and \NONRT, as a result of forming stars from pure and very diluted AGB ejecta (at the beginning and end of the simulations), respectively.
The only obvious difference is that a new peak at $Y = 0.29$ starts to form before $t = 14\Myr$ in the \RT run. Our analysis shows that this new peak is formed in the central regions. This indicates that due to a stronger ionising feedback at the centre, the pristine gas accretion (and dilution) process is somewhat weaker than it in the other regions. As a result, a slightly different dilution in the RT case results in this peak. These three helium abundances in the \INH{} simulation can qualitatively be compared with the observed values of massive clusters such as ${\rm NGC\ 2808}$ \citep{piotto2007,milone2017}, in which three populations have been found with $Y=0.248$, $Y=0.3$, and $Y =0.37$. Finally, the total SG mass formed in this high-density simulation is about $6.4\times 10^6\Msun$, which is $60$ percent of that formed in the corresponding \NONRT simulation.
%
\subsection{Role of ionising radiation in SG star formation}
\begin{figure*}
\centering
\includegraphics[width=1\linewidth]{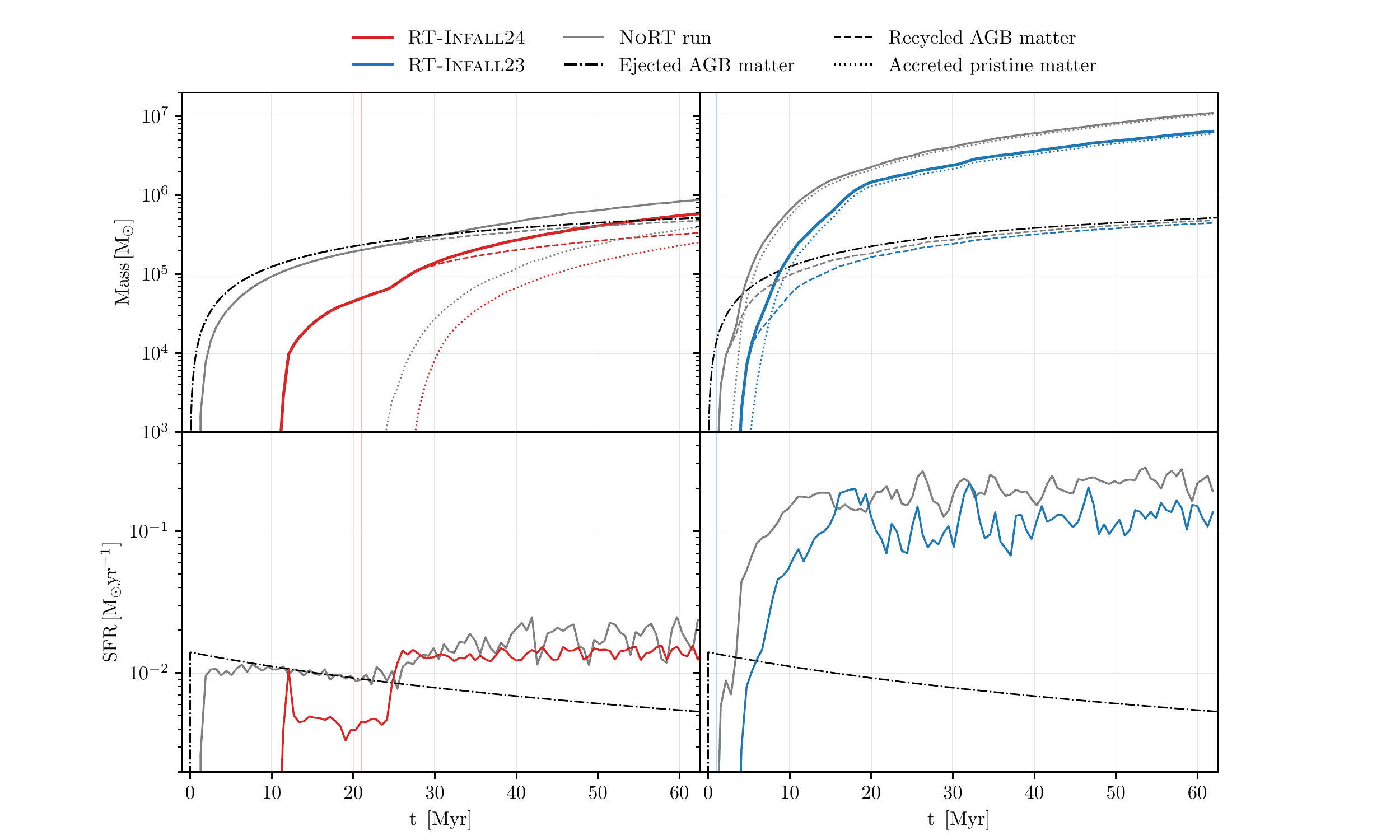}
\caption{\textbf{Top}: cumulative stellar mass of SG stars formed in our $10^7\Msun$ cluster. \textbf{Bottom}: SFR of the SG versus time for our simulated clusters. The lines are colour-coded as follows. The solid, dashed and dotted lines show the final SG stellar mass, the mass formed from AGB ejecta, and the mass formed from the pristine gas, respectively. The red and blue vertical lines mark the instance when ISM reaches the centre of the cluster in the \INL and \INH simulations, respectively. Note that our simulations starts at the onset of the AGB winds, i.e. $t_0=\tAGB$.}
\label{fig:SFR}
\end{figure*}

The top panels of \figref{fig:SFR} show the cumulative SG mass formed in our \INL and \INH simulations as a function of time. To better understand how the radiation affects the SG mass, we compare the results of our \RT simulations with the ones from the \NONRT simulations. For both cases, we also show how much of the mass of the SG stars is formed from the AGB ejecta versus the pristine gas. For comparison, the total AGB matter injected into the simulation box, computed from \equref{AGB_ej}, is also plotted.
\par In the \INL simulation (the upper left panel of the \figref{fig:SFR}), the ionising radiation delays star formation by about $10\Myr$ and also decreases the total SG mass by an order of magnitude in the first $21\Myrs$, compared to \NONRT case. Our simulations show that the cluster can retain most of its AGB ejecta in the presence of the ionising radiation even before the start of the infall, however, a large percentage of this gas is too warm  to participate in star formation. After $21\Myr$, at which point the infall reaches the centre of the cluster, the cluster immediately begins accumulating pristine gas and then the density in the central regions increases which leads to more efficient cooling. As a result, the accreted ISM makes the stellar winds cool down even more and have a larger contribution to star formation. Therefore, our results from \RT simulations emphasizes the key role of the pristine gas accretion in star formation in the context of the AGB scenario. In the \RT case, the AGB ejecta constitute about $60$ percent of the total SG mass,  almost the same as in the \NONRT case ($54\%$). Another effect of radiative heating evident in \figref{fig:SFR} is that it reduces the accumulated mass of both the AGB and the pristine gas. The total SG mass formed in the \INL is  $64\%$ of that in the \NoINL case.

\par The bottom panels of \figref{fig:SFR} show the star formation rate (SFR) in our high- and low-density simulations. As can be seen in the bottom-left panel, for the \NoINL case the SFR closely follows the rate of the injected AGB matter before the infall starts at $21\Myr$. After this time, the SFR increases and reaches a plateau with some fluctuations. This approximately constant rate of star formation can be explained by the time-independent analytic formula of the Bondi-Hoyle-Lyttleton accretion rate $\rho M^2/ v^3$ \citep[][] {Hoyle1939,Bondi1952MNRAS,Edgar2004}. Here $M$, $\rho$ and $v$ are the mass of the accretor, the density of the ambient gas and the relative velocity between the accretor and the ISM gas, respectively. In the \RT case, the SFR is smaller than that of the \NONRT case (and the ejected stellar wind rate) before $21\Myr$, and then it follows with an almost constant rate similar to the \NONRT case, albeit slightly flatter.

\par In the \INH simulation (right panels of \figref{fig:SFR}) the delay due to an earlier and denser infall is shorter. As shown in the top right panel, the SG stars are primarily formed from the accreted pristine gas, about $93\%$ in the \RT case, compared to $96\%$ in the \NONRT case. Moreover the total SG mass in the \RT run is $59$ percent of the SG mass in the \NONRT run. This panel also shows that the AGB material is well retained by the cluster. Moreover, our simulations for a $10^7\Msun$ cluster moving through the ISM with the velocity of $23\kms$, show that in environments where the ISM density is higher, more pristine gas is accumulated and then a larger fraction of stellar wind can be cooled.
Our results for a very dense medium of $10^{-22}\gocm$ exhibit similar trends and are explained in detail in Appendix \ref{sec:Appen}. The lower right panel shows that the SFR in the \RT and the \NONRT simulations also reach a plateau with a time-independent value which is approximately $10$ times that of the SFR in the low-density simulation (the lower left panel). This is also consistent with the linear dependence of the Bondi-Hoyle-Lyttleton accretion rate on the ISM density.
%
\subsection{Resolution convergence}
\begin{figure}
\centering
\includegraphics[width=1\linewidth]{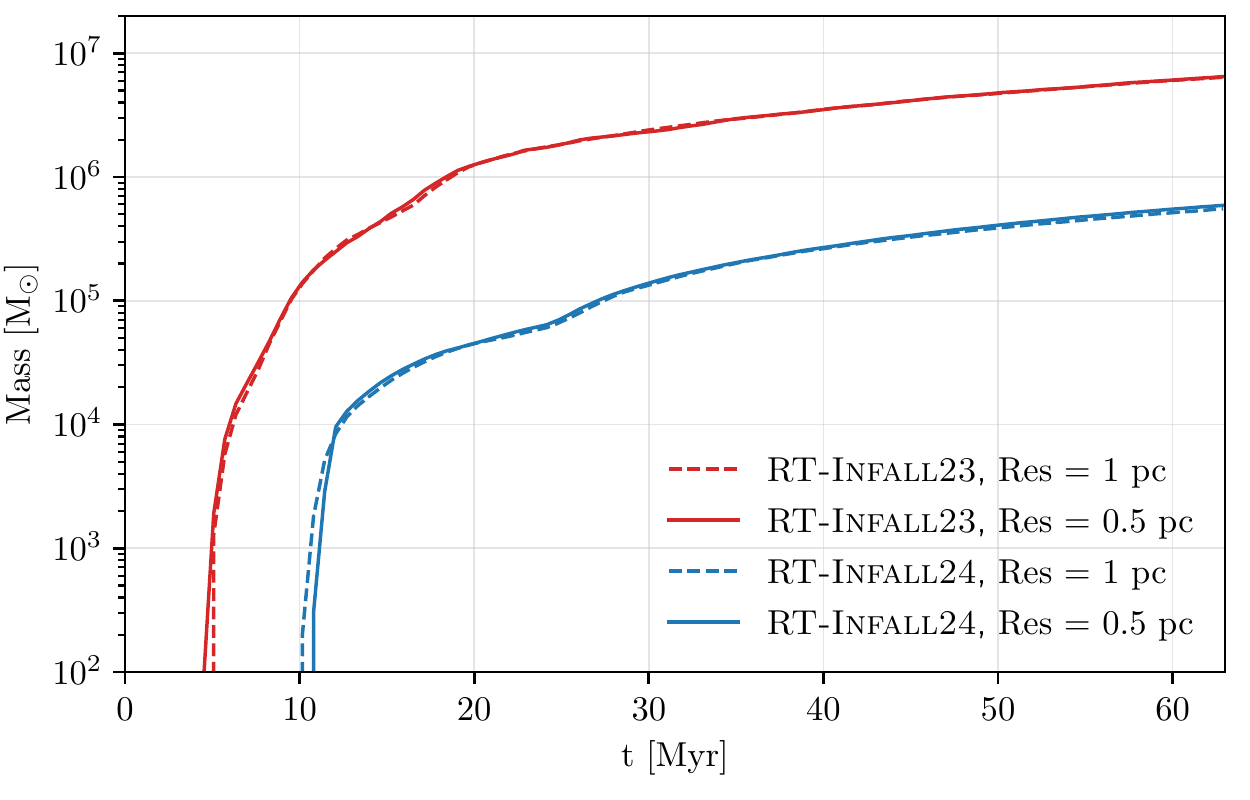}
\caption{Convergence test for cumulative SG stellar mass computed at different resolutions as a function of time. The solid and dashed lines show the results computed at the maximum resolutions of $0.5\pc$ and $1.0\pc$, respectively}
\label{fig:Res}
\end{figure}

In order to investigate the sensitivity of our results to the adopted resolution ($\Delta x_{\rm max}= 2\pc$ and $\Delta x_{\rm min}= 0.5\pc$ ) for our simulations, we performed a series of test runs. The test runs illustrate that simulations with different resolutions yield approximately the same final results, in terms of the total mass, SG profile, and He distribution. For example, \figref{fig:Res} compares the cumulative SG mass as a function of time for two resolutions ($\Delta x_{\rm min}$) of $0.5\pc$ and $1.0\pc$. These indicate that the adopted resolution is sufficient for our results to converge.
%
\section{Discussion}\label{sec:discussion}
Several studies address the formation of SG stars in GCs based on the AGB scenario. In a recent comprehensive study, using 3D hydrodynamical simulations, \citetalias{calura19} found that a very massive cluster moving through an inster-stellar medium is able to overcome the ram pressure from the ambient medium and accumulate sufficient material to form a new and massive generation of stars. In addition to the ram pressure, they also took into account self-gravity, radiative cooling, and stellar winds. Moreover, they investigated the roles of the AGB stars ejecta and the accreted pristine gas in the formation of the SG stars. 

\par Building upon \citetalias{calura19}, we have added radiative feedback from both FG and SG stars in simulations to examine how it affects the formation of SG stars. To this end, we utilized the \ramsesrt code to simulate a cluster with an initial FG mass of $10^7\Msun$ and a half-radius of $30\pc$. We placed the cluster in a homogeneous ISM, flowing towards the cluster centre at a speed of $23\kms$. We considered two densities for the ISM, namely $10^{-24}\gocm$ (\INL) and $10^{-23}\gocm$ (\INH). We started our simulations at the onset of the AGB stars winds ($t=39\Myr$) and followed the evolution of the cluster until the start of FG Type I SNe, which is $\sim100\Myr$ after the formation of the cluster. 

Our simulations include photoionisation heating and stellar winds from AGB stars. There are other sources of energy, excluded from our study because they are not expected to affect the SF history of the cluster. In particular, we ignore the effects of intermediate-mass main sequence stars \citep{naiman2018}. The exact amount of energy injected by those sources is not well determined, but their energy deposition rate can be determined as $\sim\dot{M}v_{\rm wind}^2$ \citep[e. g.][]{Weaver1977}. 
By utilising MESA models \citep{Paxton2011}, and the observed initial-final mass relation explained in \citet{naiman2018}, the mean wind velocity and mass loss of main sequence stars are estimated to be $\sim20$ times more, and $\sim10^{-4}$ times less than that of AGB stars, respectively. This translates into a wind energy ratio of $\sim0.05$, rendering the impact of intermediate-mass stars on the total energy loss negligible. Other energy sources may be present, such as planetary nebulae. However, \citet{D'Ercole2008} showed that their contribution in a massive cluster is expected to be negligible and that a total extra energy of $\sim10^{38}{\,\mathrm{erg\ s^{-1}}}$, which is $100$ times more than of AGB stars, can halt SF in such a system.

We have used the SED model of \citep{BPASS2017} for our stellar populations (see \figref{fig:SEDs}). Selecting such a population with binary stars allows us to provide the maximum possible amount of radiation in our GC. Our results with another model with lower luminosities, such as the Starburst99 \citep{S99}, show that the efficiency of radiation decreases when we assume the stellar population to consist of single stars. 
    
To examine the role of ionising radiation in the retention and cooling of He-rich stellar winds, we repeated our simulations without the infall of the pristine gas.  We found that in the \RT case, the amount of stellar winds that participate in the star formation process decreases by a factor of $\sim4$ compared to a \NONRT case without infall. Therefore increasing the density due to gas accretion within a GC results in more effective cooling and increases the SFR. This highlights the key role played by the infall of the pristine gas in the formation of SG stars in the AGB scenario. 
    
We found that the luminosity of such a $40$-Myr-old massive cluster is high enough to ionise its ICM. One might compare this result with what has been predicted by \citet{Conroy2011}. They suggested that the SG stars should have been formed several $100\Myrs$ after the FG formation, when the Lyman–Werner photon density of the FG population drops by more than three orders of magnitude. In contrast, our results including photoionisation indicate that the cooling process in a massive cluster could also be effective enough to cool the accumulated gas in the central regions and keep it neutral within a shorter time of $\sim 50$ Myr. In the case of very young clusters, it has been shown that the stellar feedback is dominated by photoionisation before the first supernova \citep{Dale2014,gavagnin2017}. As seen in \figref{fig:SEDs}, such clusters experiences very high luminosities in the first $\Myrs$ of their formation, resulting in expelling large fractions of the gas within the initial cloud and decreasing the star formation efficiency by factors $10-20$ \citep{Dale2014}. Even if such a strong feedback does not lead to a significant gas expulsion, radiative heating may not allow the gas within the cluster to cool sufficiently. This argument can be problematic for the SG formation scenarios in which the enriched stars are formed a few $\Myrs$ after the FG cluster formation. 

The chemical composition of SG stars is found to be sensitive to the time difference between the onset of the AGB ejecta and the infall of the pristine gas. However, we found that radiation did not have much effect on the He abundances of SG stars and their mass vs He abundance distribution (\figref{fig:distrib}). Moreover, the SG stars in both \RT and \NONRT simulations are more concentrated than the FG stars, in agreement with observational findings. The AGB scenario suggests that stars with higher helium abundances are formed earlier than the ones with lower He (see \figref{fig:distrib}). This can be compared with other scenarios for the formation of SG stars, for example the FRMS scenario suggests the opposite. However there is still no strong observational constraints to check this point. 
    
The majority of young (a few $\Myrs$ - $200\Myr$) massive ($\sim 10^6 \Msun$) clusters have been observed to be gas free, except for a negligible amount of ionised gas that has been detected within some of them \citep{Bastian2014,Cabrera2015}. These observations might seem to constrain the SG formation in GCs. But the point is that there is no evidence to date for the existence of the SG in such GCs, which are younger than $2$ Gyr \citep{bastian2018}. Nevertheless, the properties of the formation environment of progenitors of the GCs hosting MSPs are expected to be different from young massive star clusters. 

Our choice of simulating a very massive cluster with a mass of $10^7\Msun$ and a half-mass radius $30\pc$ might at first glance seem questionable. For example, \citet{Baumgardt2019} estimated that the initial masses of the clusters hosting MSPs are smaller. In addition, according to the flat relation between the mass and the half-mass radius for young massive clusters and GCs \citep{Portegies2010,Krumholz2019}, their half-mass radius is found to be less than $\sim3\pc$. However, we demonstrated in  \citet{Yaghoobi2022} that a $10^6\Msun$ cluster with a half-mass radius of $4\pc$ is also able to form a massive SG. Moreover a number of analytic studies \citep{lin2007,naiman2018} suggest that the cluster velocity dispersion and not the cluster mass alone, is the key parameter which ultimately determines the fate of gas accretion and the stellar wind expulsion. Therefore, as far as the formation of SG stars is concerned, our results might be compatible with a cluster of the same velocity dispersion, but a lower mass. Moreover, the exact initial conditions of GCs with MSPs is still debated and no consensus has been reached on this. Nevertheless, we have planned for a future study to investigate the role of radiative feedback in SG formation over a wider range of initial cluster masses and half-mass radii, including lower cluster masses.
%
\section{Conclusions}\label{sec:conclusion}
We investigate whether stellar photoionisation feedback can influence SG formation in a very massive GC in the context of the AGB scenario. We simulate SG formation in two media of different densities and use a stellar population model that includes binary stars. The main outcomes of our study can be summarized as follows.

\begin{itemize}
    \item In a cluster with a mass of $10^7\Msun$ and an age of $40\Myr$, long after FG Type II SNe have subsided, photoionisation does not lead to a significant gas expulsion. As a result, the ejecta of the AGB stars and the accreted pristine gas can be retained in the central regions of the cluster. Our results indicate that the cooling process in a massive cluster could also be effective enough to cool the accumulated gas in the central regions and keep it neutral within a time of $\sim 50\Myr$.
    
    \item The inclusion of photoionisation delays the formation of SG stars. It also leads to a modest decrease in the total SG mass. The extent of this effect depends on the ISM density so that a denser medium, decreases the delay in star formation. 
        
     \item In the case of the low-density simulation (\INL), radiative heating does not allow any tail to be formed behind the cluster. It, however, makes the distribution of SG stars elongated. The gas is everywhere ionised until $\sim 10\Myr$ after injecting the AGB ejecta. After that, only at the very central regions does the gas become cold enough to turn into stars. The final mass of SG stars in this model is about $6\times10^5\Msun$, approximately $67$ percent of that in the \NoINL{}. 
     
     \item In the high-density case (\INH), the stellar luminosity is not sufficient to ionise all the gas in the box due to a denser infall. However, more accumulated gas within the cluster leads to a more efficient cooling in the central region, and thus star formation occurs earlier than the low density model. Finally, the final SG mass formed in this case is about $6.4\times 10^6\Msun$ ($60$ percent of that in the \NoINH simulation). Moreover, three sub-populations with different helium abundances, in a good agreement with observations of massive clusters, are found. In addition, a strong tail is formed behind the cluster in this model.
     
     \item In all the simulations, we find that the radiation pressure hardly affects the star formation within the cluster, i.e. all the effect of radiation is from photoionisation heating. The radiative effects of the SG population are also found to be negligible in the present work.

\end{itemize}

\section*{Acknowledgements}
AY is grateful to the Centre de Recherche Astrophysique (CRAL) for hospitality during her visit.  FC acknowledges support from PRIN INAF 1.05.01.85.01 and INAF Main-Stream 1.05.01.86.31.
We acknowledge support and computational resources from the Common Computing Facility (CCF) of the LABEX Lyon Institute of Origins (ANR-10-LABX-66) and the PSMN (Pôle Scientifique de Modélisation Numérique) of the ENS de Lyon. We thank the referee, Richard Wünsch, for a constructive review.

\section*{DATA AVAILABILITY}
The data that support the findings of this study are available from the corresponding author upon reasonable request.
%
\bibliographystyle{mnras}
\input{main.bbl}

\begin{thebibliography}{}
\makeatletter
\relax
\def\mn@urlcharsother{\let\do\@makeother \do\$\do\&\do\#\do\^\do\_\do\%\do\~}
\def\mn@doi{\begingroup\mn@urlcharsother \@ifnextchar [ {\mn@doi@}
  {\mn@doi@[]}}
\def\mn@doi@[#1]#2{\def\@tempa{#1}\ifx\@tempa\@empty \href
  {http://dx.doi.org/#2} {doi:#2}\else \href {http://dx.doi.org/#2} {#1}\fi
  \endgroup}
\def\mn@eprint#1#2{\mn@eprint@#1:#2::\@nil}
\def\mn@eprint@arXiv#1{\href {http://arxiv.org/abs/#1} {{\tt arXiv:#1}}}
\def\mn@eprint@dblp#1{\href {http://dblp.uni-trier.de/rec/bibtex/#1.xml}
  {dblp:#1}}
\def\mn@eprint@#1:#2:#3:#4\@nil{\def\@tempa {#1}\def\@tempb {#2}\def\@tempc
  {#3}\ifx \@tempc \@empty \let \@tempc \@tempb \let \@tempb \@tempa \fi \ifx
  \@tempb \@empty \def\@tempb {arXiv}\fi \@ifundefined
  {mn@eprint@\@tempb}{\@tempb:\@tempc}{\expandafter \expandafter \csname
  mn@eprint@\@tempb\endcsname \expandafter{\@tempc}}}

\bibitem[\protect\citeauthoryear{{Bastian} \& {Lardo}}{{Bastian} \&
  {Lardo}}{2018}]{bastian2018}
{Bastian} N.,  {Lardo} C.,  2018, \mn@doi [\araa]
  {10.1146/annurev-astro-081817-051839}, \href
  {https://ui.adsabs.harvard.edu/abs/2018ARA&A..56...83B} {56, 83}

\bibitem[\protect\citeauthoryear{{Bastian}, {Lamers}, {de Mink}, {Longmore},
  {Goodwin}  \& {Gieles}}{{Bastian} et~al.}{2013}]{bastian2013}
{Bastian} N.,  {Lamers} H.~J.~G.~L.~M.,  {de Mink} S.~E.,  {Longmore} S.~N.,
  {Goodwin} S.~P.,   {Gieles} M.,  2013, \mn@doi [\mnras]
  {10.1093/mnras/stt1745}, \href
  {https://ui.adsabs.harvard.edu/abs/2013MNRAS.436.2398B} {436, 2398}

\bibitem[\protect\citeauthoryear{{Bastian}, {Hollyhead}  \&
  {Cabrera-Ziri}}{{Bastian} et~al.}{2014}]{Bastian2014}
{Bastian} N.,  {Hollyhead} K.,   {Cabrera-Ziri} I.,  2014, \mn@doi [\mnras]
  {10.1093/mnras/stu1775}, \href
  {https://ui.adsabs.harvard.edu/abs/2014MNRAS.445..378B} {445, 378}

\bibitem[\protect\citeauthoryear{{Bastian}, {Cabrera-Ziri}  \&
  {Salaris}}{{Bastian} et~al.}{2015}]{bastian2015}
{Bastian} N.,  {Cabrera-Ziri} I.,   {Salaris} M.,  2015, \mn@doi [\mnras]
  {10.1093/mnras/stv543}, \href
  {https://ui.adsabs.harvard.edu/abs/2015MNRAS.449.3333B} {449, 3333}

\bibitem[\protect\citeauthoryear{{Baumgardt}, {Hilker}, {Sollima}  \&
  {Bellini}}{{Baumgardt} et~al.}{2019}]{Baumgardt2019}
{Baumgardt} H.,  {Hilker} M.,  {Sollima} A.,   {Bellini} A.,  2019, \mn@doi
  [\mnras] {10.1093/mnras/sty2997}, \href
  {https://ui.adsabs.harvard.edu/abs/2019MNRAS.482.5138B} {482, 5138}

\bibitem[\protect\citeauthoryear{{Bekki}}{{Bekki}}{2017}]{Bekki2017}
{Bekki} K.,  2017, \mn@doi [\mnras] {10.1093/mnras/stx110}, \href
  {https://ui.adsabs.harvard.edu/abs/2017MNRAS.467.1857B} {467, 1857}

\bibitem[\protect\citeauthoryear{{Bekki}}{{Bekki}}{2019}]{bekki2019}
{Bekki} K.,  2019, \mn@doi [\mnras] {10.1093/mnras/stz999}, \href
  {https://ui.adsabs.harvard.edu/abs/2019MNRAS.486.2570B} {486, 2570}

\bibitem[\protect\citeauthoryear{{Bekki} \& {Freeman}}{{Bekki} \&
  {Freeman}}{2003}]{Bekki2003}
{Bekki} K.,  {Freeman} K.~C.,  2003, \mn@doi [\mnras]
  {10.1046/j.1365-2966.2003.07275.x}, \href
  {https://ui.adsabs.harvard.edu/abs/2003MNRAS.346L..11B} {346, L11}

\bibitem[\protect\citeauthoryear{{Bekki}, {Je{\v{r}}{\'a}bkov{\'a}}  \&
  {Kroupa}}{{Bekki} et~al.}{2017}]{BekkiKroupa2017}
{Bekki} K.,  {Je{\v{r}}{\'a}bkov{\'a}} T.,   {Kroupa} P.,  2017, \mn@doi
  [\mnras] {10.1093/mnras/stx1609}, \href
  {https://ui.adsabs.harvard.edu/abs/2017MNRAS.471.2242B} {471, 2242}

\bibitem[\protect\citeauthoryear{{Bondi}}{{Bondi}}{1952}]{Bondi1952MNRAS}
{Bondi} H.,  1952, \mn@doi [\mnras] {10.1093/mnras/112.2.195}, \href
  {https://ui.adsabs.harvard.edu/abs/1952MNRAS.112..195B} {112, 195}

\bibitem[\protect\citeauthoryear{{Bragaglia}, {Carretta}, {Gratton}, {D'Orazi},
  {Lucatello}  \& {Sneden}}{{Bragaglia} et~al.}{2013}]{Bragaglia2013}
{Bragaglia} A.,  {Carretta} E.,  {Gratton} R.,  {D'Orazi} V.,  {Lucatello} S.,
   {Sneden} C.,  2013, \memsai, \href
  {https://ui.adsabs.harvard.edu/abs/2013MmSAI..84...24B} {84, 24}

\bibitem[\protect\citeauthoryear{{Bragaglia}, {Sneden}, {Carretta}, {Gratton},
  {Lucatello}, {Bernath}, {Brooke}  \& {Ram}}{{Bragaglia}
  et~al.}{2014}]{Bragaglia2014}
{Bragaglia} A.,  {Sneden} C.,  {Carretta} E.,  {Gratton} R.~G.,  {Lucatello}
  S.,  {Bernath} P.~F.,  {Brooke} J. S.~A.,   {Ram} R.~S.,  2014, \mn@doi
  [\apj] {10.1088/0004-637X/796/1/68}, \href
  {https://ui.adsabs.harvard.edu/abs/2014ApJ...796...68B} {796, 68}

\bibitem[\protect\citeauthoryear{{Cabrera-Ziri} et~al.,}{{Cabrera-Ziri}
  et~al.}{2015}]{Cabrera2015}
{Cabrera-Ziri} I.,  et~al., 2015, \mn@doi [\mnras] {10.1093/mnras/stv163},
  \href {https://ui.adsabs.harvard.edu/abs/2015MNRAS.448.2224C} {448, 2224}

\bibitem[\protect\citeauthoryear{{Calura}, {Few}, {Romano}  \&
  {D'Ercole}}{{Calura} et~al.}{2015}]{calura2015}
{Calura} F.,  {Few} C.~G.,  {Romano} D.,   {D'Ercole} A.,  2015, \mn@doi
  [\apjl] {10.1088/2041-8205/814/1/L14}, \href
  {https://ui.adsabs.harvard.edu/abs/2015ApJ...814L..14C} {814, L14}

\bibitem[\protect\citeauthoryear{{Calura}, {D'Ercole}, {Vesperini}, {Vanzella}
  \& {Sollima}}{{Calura} et~al.}{2019}]{calura19}
{Calura} F.,  {D'Ercole} A.,  {Vesperini} E.,  {Vanzella} E.,   {Sollima} A.,
  2019, \mn@doi [\mnras] {10.1093/mnras/stz2055}, \href
  {https://ui.adsabs.harvard.edu/abs/2019MNRAS.489.3269C} {489, 3269}

\bibitem[\protect\citeauthoryear{{Carretta} et~al.,}{{Carretta}
  et~al.}{2009}]{carretta2009}
{Carretta} E.,  et~al., 2009, \mn@doi [\aap] {10.1051/0004-6361/200912096},
  \href {https://ui.adsabs.harvard.edu/abs/2009A&A...505..117C} {505, 117}

\bibitem[\protect\citeauthoryear{{Carretta}, {Bragaglia}, {Gratton},
  {Recio-Blanco}, {Lucatello}, {D'Orazi}  \& {Cassisi}}{{Carretta}
  et~al.}{2010}]{carretta2010}
{Carretta} E.,  {Bragaglia} A.,  {Gratton} R.~G.,  {Recio-Blanco} A.,
  {Lucatello} S.,  {D'Orazi} V.,   {Cassisi} S.,  2010, \mn@doi [\aap]
  {10.1051/0004-6361/200913451}, \href
  {https://ui.adsabs.harvard.edu/abs/2010A&A...516A..55C} {516, A55}

\bibitem[\protect\citeauthoryear{{Chantereau}, {Biernacki}, {Martig},
  {Bastian}, {Salaris}  \& {Teyssier}}{{Chantereau}
  et~al.}{2020}]{chantereau2020}
{Chantereau} W.,  {Biernacki} P.,  {Martig} M.,  {Bastian} N.,  {Salaris} M.,
  {Teyssier} R.,  2020, \mn@doi [\mnras] {10.1093/mnras/staa371}, \href
  {https://ui.adsabs.harvard.edu/abs/2020MNRAS.493.1306C} {493, 1306}

\bibitem[\protect\citeauthoryear{{Ciotti}, {D'Ercole}, {Pellegrini}  \&
  {Renzini}}{{Ciotti} et~al.}{1991}]{ciotti1991}
{Ciotti} L.,  {D'Ercole} A.,  {Pellegrini} S.,   {Renzini} A.,  1991, \mn@doi
  [\apj] {10.1086/170289}, \href
  {https://ui.adsabs.harvard.edu/abs/1991ApJ...376..380C} {376, 380}

\bibitem[\protect\citeauthoryear{{Conroy} \& {Spergel}}{{Conroy} \&
  {Spergel}}{2011}]{Conroy2011}
{Conroy} C.,  {Spergel} D.~N.,  2011, \mn@doi [\apj]
  {10.1088/0004-637X/726/1/36}, \href
  {https://ui.adsabs.harvard.edu/abs/2011ApJ...726...36C} {726, 36}

\bibitem[\protect\citeauthoryear{{Conroy}, {Gunn}  \& {White}}{{Conroy}
  et~al.}{2009}]{Conroy2009}
{Conroy} C.,  {Gunn} J.~E.,   {White} M.,  2009, \mn@doi [\apj]
  {10.1088/0004-637X/699/1/486}, \href
  {https://ui.adsabs.harvard.edu/abs/2009ApJ...699..486C} {699, 486}

\bibitem[\protect\citeauthoryear{{D'Ercole}, {Vesperini}, {D'Antona},
  {McMillan}  \& {Recchi}}{{D'Ercole} et~al.}{2008}]{D'Ercole2008}
{D'Ercole} A.,  {Vesperini} E.,  {D'Antona} F.,  {McMillan} S. L.~W.,
  {Recchi} S.,  2008, \mn@doi [\mnras] {10.1111/j.1365-2966.2008.13915.x},
  \href {https://ui.adsabs.harvard.edu/abs/2008MNRAS.391..825D} {391, 825}

\bibitem[\protect\citeauthoryear{{D'Ercole}, {D'Antona}  \&
  {Vesperini}}{{D'Ercole} et~al.}{2016}]{D'Ercole2016}
{D'Ercole} A.,  {D'Antona} F.,   {Vesperini} E.,  2016, \mn@doi [\mnras]
  {10.1093/mnras/stw1583}, \href
  {https://ui.adsabs.harvard.edu/abs/2016MNRAS.461.4088D} {461, 4088}

\bibitem[\protect\citeauthoryear{{Dale}, {Ngoumou}, {Ercolano}  \&
  {Bonnell}}{{Dale} et~al.}{2014}]{Dale2014}
{Dale} J.~E.,  {Ngoumou} J.,  {Ercolano} B.,   {Bonnell} I.~A.,  2014, \mn@doi
  [\mnras] {10.1093/mnras/stu816}, \href
  {https://ui.adsabs.harvard.edu/abs/2014MNRAS.442..694D} {442, 694}

\bibitem[\protect\citeauthoryear{{Decressin}, {Meynet}, {Charbonnel},
  {Prantzos}  \& {Ekstr{\"o}m}}{{Decressin} et~al.}{2007a}]{decressin2007}
{Decressin} T.,  {Meynet} G.,  {Charbonnel} C.,  {Prantzos} N.,   {Ekstr{\"o}m}
  S.,  2007a, \mn@doi [\aap] {10.1051/0004-6361:20066013}, \href
  {https://ui.adsabs.harvard.edu/abs/2007A&A...464.1029D} {464, 1029}

\bibitem[\protect\citeauthoryear{{Decressin}, {Charbonnel}  \&
  {Meynet}}{{Decressin} et~al.}{2007b}]{DecCharbMey2007}
{Decressin} T.,  {Charbonnel} C.,   {Meynet} G.,  2007b, \mn@doi [\aap]
  {10.1051/0004-6361:20078425}, \href
  {https://ui.adsabs.harvard.edu/abs/2007A&A...475..859D} {475, 859}

\bibitem[\protect\citeauthoryear{{Denissenkov} \& {Hartwick}}{{Denissenkov} \&
  {Hartwick}}{2014}]{Denissenkov2014}
{Denissenkov} P.~A.,  {Hartwick} F.~D.~A.,  2014, \mn@doi [\mnras]
  {10.1093/mnrasl/slt133}, \href
  {https://ui.adsabs.harvard.edu/abs/2014MNRAS.437L..21D} {437, L21}

\bibitem[\protect\citeauthoryear{{Denissenkov}, {VandenBerg}, {Hartwick},
  {Herwig}, {Weiss}  \& {Paxton}}{{Denissenkov} et~al.}{2015}]{Denissenkov2015}
{Denissenkov} P.~A.,  {VandenBerg} D.~A.,  {Hartwick} F.~D.~A.,  {Herwig} F.,
  {Weiss} A.,   {Paxton} B.,  2015, \mn@doi [\mnras] {10.1093/mnras/stv211},
  \href {https://ui.adsabs.harvard.edu/abs/2015MNRAS.448.3314D} {448, 3314}

\bibitem[\protect\citeauthoryear{{Edgar}}{{Edgar}}{2004}]{Edgar2004}
{Edgar} R.,  2004, \mn@doi [\nar] {10.1016/j.newar.2004.06.001}, \href
  {https://ui.adsabs.harvard.edu/abs/2004NewAR..48..843E} {48, 843}

\bibitem[\protect\citeauthoryear{{Eldridge}, {Stanway}, {Xiao}, {McClelland},
  {Taylor}, {Ng}, {Greis}  \& {Bray}}{{Eldridge} et~al.}{2017}]{BPASS2017}
{Eldridge} J.~J.,  {Stanway} E.~R.,  {Xiao} L.,  {McClelland} L.~A.~S.,
  {Taylor} G.,  {Ng} M.,  {Greis} S.~M.~L.,   {Bray} J.~C.,  2017, \mn@doi
  [\pasa] {10.1017/pasa.2017.51}, \href
  {https://ui.adsabs.harvard.edu/abs/2017PASA...34...58E} {34, e058}

\bibitem[\protect\citeauthoryear{{Frelijj}, {Villanova}, {Mu{\~n}oz}  \&
  {Fern{\'a}ndez-Trincado}}{{Frelijj} et~al.}{2021}]{Frelijj2021}
{Frelijj} H.,  {Villanova} S.,  {Mu{\~n}oz} C.,   {Fern{\'a}ndez-Trincado}
  J.~G.,  2021, \mn@doi [\mnras] {10.1093/mnras/stab443}, \href
  {https://ui.adsabs.harvard.edu/abs/2021MNRAS.503..867F} {503, 867}

\bibitem[\protect\citeauthoryear{{Gavagnin}, {Bleuler}, {Rosdahl}  \&
  {Teyssier}}{{Gavagnin} et~al.}{2017}]{gavagnin2017}
{Gavagnin} E.,  {Bleuler} A.,  {Rosdahl} J.,   {Teyssier} R.,  2017, \mn@doi
  [\mnras] {10.1093/mnras/stx2222}, \href
  {https://ui.adsabs.harvard.edu/abs/2017MNRAS.472.4155G} {472, 4155}

\bibitem[\protect\citeauthoryear{{Gilbert} \& {Graham}}{{Gilbert} \&
  {Graham}}{2007}]{Gilbert2007}
{Gilbert} A.~M.,  {Graham} J.~R.,  2007, \mn@doi [\apj] {10.1086/520910}, \href
  {https://ui.adsabs.harvard.edu/abs/2007ApJ...668..168G} {668, 168}

\bibitem[\protect\citeauthoryear{{Gratton} et~al.,}{{Gratton}
  et~al.}{2013}]{gratton2013}
{Gratton} R.~G.,  et~al., 2013, \mn@doi [\aap] {10.1051/0004-6361/201219976},
  \href {https://ui.adsabs.harvard.edu/abs/2013A&A...549A..41G} {549, A41}

\bibitem[\protect\citeauthoryear{{Hoyle} \& {Lyttleton}}{{Hoyle} \&
  {Lyttleton}}{1939}]{Hoyle1939}
{Hoyle} F.,  {Lyttleton} R.~A.,  1939, \mn@doi [Proceedings of the Cambridge
  Philosophical Society] {10.1017/S0305004100021150}, \href
  {https://ui.adsabs.harvard.edu/abs/1939PCPS...35..405H} {35, 405}

\bibitem[\protect\citeauthoryear{{Jacquet} \& {Krumholz}}{{Jacquet} \&
  {Krumholz}}{2011}]{Jacquet2011}
{Jacquet} E.,  {Krumholz} M.~R.,  2011, \mn@doi [\apj]
  {10.1088/0004-637X/730/2/116}, \href
  {https://ui.adsabs.harvard.edu/abs/2011ApJ...730..116J} {730, 116}

\bibitem[\protect\citeauthoryear{{Je{\v{r}}{\'a}bkov{\'a}}, {Hasani Zonoozi},
  {Kroupa}, {Beccari}, {Yan}, {Vazdekis}  \& {Zhang}}{{Je{\v{r}}{\'a}bkov{\'a}}
  et~al.}{2018}]{Tereza2018}
{Je{\v{r}}{\'a}bkov{\'a}} T.,  {Hasani Zonoozi} A.,  {Kroupa} P.,  {Beccari}
  G.,  {Yan} Z.,  {Vazdekis} A.,   {Zhang} Z.~Y.,  2018, \mn@doi [\aap]
  {10.1051/0004-6361/201833055}, \href
  {https://ui.adsabs.harvard.edu/abs/2018A&A...620A..39J} {620, A39}

\bibitem[\protect\citeauthoryear{{Khalaj} \& {Baumgardt}}{{Khalaj} \&
  {Baumgardt}}{2015}]{Khalaj2015}
{Khalaj} P.,  {Baumgardt} H.,  2015, \mn@doi [\mnras] {10.1093/mnras/stv1356},
  \href {https://ui.adsabs.harvard.edu/abs/2015MNRAS.452..924K} {452, 924}

\bibitem[\protect\citeauthoryear{{Khalaj} \& {Baumgardt}}{{Khalaj} \&
  {Baumgardt}}{2016}]{Khalaj2016}
{Khalaj} P.,  {Baumgardt} H.,  2016, \mn@doi [\mnras] {10.1093/mnras/stv2944},
  \href {https://ui.adsabs.harvard.edu/abs/2016MNRAS.457..479K} {457, 479}

\bibitem[\protect\citeauthoryear{{Krause}, {Charbonnel}, {Decressin}, {Meynet}
  \& {Prantzos}}{{Krause} et~al.}{2013}]{krause2013}
{Krause} M.,  {Charbonnel} C.,  {Decressin} T.,  {Meynet} G.,   {Prantzos} N.,
  2013, \mn@doi [\aap] {10.1051/0004-6361/201220694}, \href
  {https://ui.adsabs.harvard.edu/abs/2013A&A...552A.121K} {552, A121}

\bibitem[\protect\citeauthoryear{{Kroupa}}{{Kroupa}}{2001}]{kroupa2001}
{Kroupa} P.,  2001, \mn@doi [\mnras] {10.1046/j.1365-8711.2001.04022.x}, \href
  {https://ui.adsabs.harvard.edu/abs/2001MNRAS.322..231K} {322, 231}

\bibitem[\protect\citeauthoryear{{Kroupa}, {Je{\v{r}}{\'a}bkov{\'a}},
  {Dinnbier}, {Beccari}  \& {Yan}}{{Kroupa} et~al.}{2018}]{Kroupa2018}
{Kroupa} P.,  {Je{\v{r}}{\'a}bkov{\'a}} T.,  {Dinnbier} F.,  {Beccari} G.,
  {Yan} Z.,  2018, \mn@doi [\aap] {10.1051/0004-6361/201732151}, \href
  {https://ui.adsabs.harvard.edu/abs/2018A&A...612A..74K} {612, A74}

\bibitem[\protect\citeauthoryear{{Krumholz}, {McKee}  \&
  {Bland-Hawthorn}}{{Krumholz} et~al.}{2019}]{Krumholz2019}
{Krumholz} M.~R.,  {McKee} C.~F.,   {Bland-Hawthorn} J.,  2019, \mn@doi [\araa]
  {10.1146/annurev-astro-091918-104430}, \href
  {https://ui.adsabs.harvard.edu/abs/2019ARA&A..57..227K} {57, 227}

\bibitem[\protect\citeauthoryear{{Lacchin}, {Calura}  \& {Vesperini}}{{Lacchin}
  et~al.}{2021}]{lacchin21}
{Lacchin} E.,  {Calura} F.,   {Vesperini} E.,  2021, \mn@doi [\mnras]
  {10.1093/mnras/stab2061}, \href
  {https://ui.adsabs.harvard.edu/abs/2021MNRAS.506.5951L} {506, 5951}

\bibitem[\protect\citeauthoryear{{Lagioia}, {Milone}, {Marino}, {Cordoni}  \&
  {Tailo}}{{Lagioia} et~al.}{2019}]{Lagioia2019}
{Lagioia} E.~P.,  {Milone} A.~P.,  {Marino} A.~F.,  {Cordoni} G.,   {Tailo} M.,
   2019, \mn@doi [\aj] {10.3847/1538-3881/ab45f2}, \href
  {https://ui.adsabs.harvard.edu/abs/2019AJ....158..202L} {158, 202}

\bibitem[\protect\citeauthoryear{{Leitherer} et~al.,}{{Leitherer}
  et~al.}{1999}]{S99}
{Leitherer} C.,  et~al., 1999, \mn@doi [\apjs] {10.1086/313233}, \href
  {https://ui.adsabs.harvard.edu/abs/1999ApJS..123....3L} {123, 3}

\bibitem[\protect\citeauthoryear{{Lin} \& {Murray}}{{Lin} \&
  {Murray}}{2007}]{lin2007}
{Lin} D. N.~C.,  {Murray} S.~D.,  2007, \mn@doi [\apj] {10.1086/515387}, \href
  {https://ui.adsabs.harvard.edu/abs/2007ApJ...661..779L} {661, 779}

\bibitem[\protect\citeauthoryear{{Marcolini}, {Brighenti}  \&
  {D'Ercole}}{{Marcolini} et~al.}{2003}]{marcolini2003}
{Marcolini} A.,  {Brighenti} F.,   {D'Ercole} A.,  2003, \mn@doi [\mnras]
  {10.1046/j.1365-2966.2003.07054.x}, \href
  {https://ui.adsabs.harvard.edu/abs/2003MNRAS.345.1329M} {345, 1329}

\bibitem[\protect\citeauthoryear{{Milone} et~al.,}{{Milone}
  et~al.}{2017}]{milone2017}
{Milone} A.~P.,  et~al., 2017, \mn@doi [\mnras] {10.1093/mnras/stw2531}, \href
  {https://ui.adsabs.harvard.edu/abs/2017MNRAS.464.3636M} {464, 3636}

\bibitem[\protect\citeauthoryear{{Milone} et~al.,}{{Milone}
  et~al.}{2020}]{milone2020}
{Milone} A.~P.,  et~al., 2020, \mn@doi [\mnras] {10.1093/mnras/stz2999}, \href
  {https://ui.adsabs.harvard.edu/abs/2020MNRAS.491..515M} {491, 515}

\bibitem[\protect\citeauthoryear{{Minniti}, {Geisler}, {Peterson}  \&
  {Claria}}{{Minniti} et~al.}{1993}]{minniti1993}
{Minniti} D.,  {Geisler} D.,  {Peterson} R.~C.,   {Claria} J.~J.,  1993,
  \mn@doi [\apj] {10.1086/173024}, \href
  {https://ui.adsabs.harvard.edu/abs/1993ApJ...413..548M} {413, 548}

\bibitem[\protect\citeauthoryear{{Naiman}, {Ramirez-Ruiz}  \& {Lin}}{{Naiman}
  et~al.}{2011}]{naiman2011}
{Naiman} J.~P.,  {Ramirez-Ruiz} E.,   {Lin} D.~N.~C.,  2011, \mn@doi [\apj]
  {10.1088/0004-637X/735/1/25}, \href
  {https://ui.adsabs.harvard.edu/abs/2011ApJ...735...25N} {735, 25}

\bibitem[\protect\citeauthoryear{{Naiman}, {Ramirez-Ruiz}  \& {Lin}}{{Naiman}
  et~al.}{2018}]{naiman2018}
{Naiman} J.~P.,  {Ramirez-Ruiz} E.,   {Lin} D.~N.~C.,  2018, \mn@doi [\mnras]
  {10.1093/mnras/sty1198}, \href
  {https://ui.adsabs.harvard.edu/abs/2018MNRAS.478.2794N} {478, 2794}

\bibitem[\protect\citeauthoryear{{Pancino}, {Carrera}, {Rossetti}  \&
  {Gallart}}{{Pancino} et~al.}{2010}]{Pancino2010}
{Pancino} E.,  {Carrera} R.,  {Rossetti} E.,   {Gallart} C.,  2010, \mn@doi
  [\aap] {10.1051/0004-6361/200912965}, \href
  {https://ui.adsabs.harvard.edu/abs/2010A&A...511A..56P} {511, A56}

\bibitem[\protect\citeauthoryear{{Paxton}, {Bildsten}, {Dotter}, {Herwig},
  {Lesaffre}  \& {Timmes}}{{Paxton} et~al.}{2011}]{Paxton2011}
{Paxton} B.,  {Bildsten} L.,  {Dotter} A.,  {Herwig} F.,  {Lesaffre} P.,
  {Timmes} F.,  2011, \mn@doi [\apjs] {10.1088/0067-0049/192/1/3}, \href
  {https://ui.adsabs.harvard.edu/abs/2011ApJS..192....3P} {192, 3}

\bibitem[\protect\citeauthoryear{{Piotto} et~al.,}{{Piotto}
  et~al.}{2007}]{piotto2007}
{Piotto} G.,  et~al., 2007, \mn@doi [\apjl] {10.1086/518503}, \href
  {https://ui.adsabs.harvard.edu/abs/2007ApJ...661L..53P} {661, L53}

\bibitem[\protect\citeauthoryear{{Plummer}}{{Plummer}}{1911}]{plummer1911}
{Plummer} H.~C.,  1911, \mn@doi [\mnras] {10.1093/mnras/71.5.460}, \href
  {https://ui.adsabs.harvard.edu/abs/1911MNRAS..71..460P} {71, 460}

\bibitem[\protect\citeauthoryear{{Portegies Zwart}, {McMillan}  \&
  {Gieles}}{{Portegies Zwart} et~al.}{2010}]{Portegies2010}
{Portegies Zwart} S.~F.,  {McMillan} S. L.~W.,   {Gieles} M.,  2010, \mn@doi
  [\araa] {10.1146/annurev-astro-081309-130834}, \href
  {https://ui.adsabs.harvard.edu/abs/2010ARA&A..48..431P} {48, 431}

\bibitem[\protect\citeauthoryear{{Rasera} \& {Teyssier}}{{Rasera} \&
  {Teyssier}}{2006}]{Rasera2006}
{Rasera} Y.,  {Teyssier} R.,  2006, \mn@doi [\aap]
  {10.1051/0004-6361:20053116}, \href
  {https://ui.adsabs.harvard.edu/abs/2006A&A...445....1R} {445, 1}

\bibitem[\protect\citeauthoryear{{Renzini} et~al.,}{{Renzini}
  et~al.}{2015}]{renzini2015}
{Renzini} A.,  et~al., 2015, \mn@doi [\mnras] {10.1093/mnras/stv2268}, \href
  {https://ui.adsabs.harvard.edu/abs/2015MNRAS.454.4197R} {454, 4197}

\bibitem[\protect\citeauthoryear{{Rosdahl} \& {Teyssier}}{{Rosdahl} \&
  {Teyssier}}{2015}]{Rosdahl2015}
{Rosdahl} J.,  {Teyssier} R.,  2015, \mn@doi [\mnras] {10.1093/mnras/stv567},
  \href {https://ui.adsabs.harvard.edu/abs/2015MNRAS.449.4380R} {449, 4380}

\bibitem[\protect\citeauthoryear{{Rosdahl}, {Blaizot}, {Aubert}, {Stranex}  \&
  {Teyssier}}{{Rosdahl} et~al.}{2013}]{rosdahl2013}
{Rosdahl} J.,  {Blaizot} J.,  {Aubert} D.,  {Stranex} T.,   {Teyssier} R.,
  2013, \mn@doi [\mnras] {10.1093/mnras/stt1722}, \href
  {https://ui.adsabs.harvard.edu/abs/2013MNRAS.436.2188R} {436, 2188}

\bibitem[\protect\citeauthoryear{{Rosen} \& {Bregman}}{{Rosen} \&
  {Bregman}}{1995}]{Rosen1995ApJ}
{Rosen} A.,  {Bregman} J.~N.,  1995, \mn@doi [\apj] {10.1086/175303}, \href
  {https://ui.adsabs.harvard.edu/abs/1995ApJ...440..634R} {440, 634}

\bibitem[\protect\citeauthoryear{{Schmidt}}{{Schmidt}}{1959}]{Schmidt1959}
{Schmidt} M.,  1959, \mn@doi [\apj] {10.1086/146614}, \href
  {https://ui.adsabs.harvard.edu/abs/1959ApJ...129..243S} {129, 243}

\bibitem[\protect\citeauthoryear{{Sutherland} \& {Dopita}}{{Sutherland} \&
  {Dopita}}{1993}]{sutherland1993cooling}
{Sutherland} R.~S.,  {Dopita} M.~A.,  1993, \mn@doi [\apjs] {10.1086/191823},
  \href {https://ui.adsabs.harvard.edu/abs/1993ApJS...88..253S} {88, 253}

\bibitem[\protect\citeauthoryear{{Sz{\'e}csi}, {Mackey}  \&
  {Langer}}{{Sz{\'e}csi} et~al.}{2018}]{Szesi2018}
{Sz{\'e}csi} D.,  {Mackey} J.,   {Langer} N.,  2018, \mn@doi [\aap]
  {10.1051/0004-6361/201731500}, \href
  {https://ui.adsabs.harvard.edu/abs/2018A&A...612A..55S} {612, A55}

\bibitem[\protect\citeauthoryear{{Teyssier}}{{Teyssier}}{2002}]{teyssier2002}
{Teyssier} R.,  2002, \mn@doi [\aap] {10.1051/0004-6361:20011817}, \href
  {https://ui.adsabs.harvard.edu/abs/2002A&A...385..337T} {385, 337}

\bibitem[\protect\citeauthoryear{{Ventura} \& {D'Antona}}{{Ventura} \&
  {D'Antona}}{2011}]{ventura2011}
{Ventura} P.,  {D'Antona} F.,  2011, \mn@doi [\mnras]
  {10.1111/j.1365-2966.2010.17651.x}, \href
  {https://ui.adsabs.harvard.edu/abs/2011MNRAS.410.2760V} {410, 2760}

\bibitem[\protect\citeauthoryear{{Villanova}, {Geisler}, {Carraro}, {Moni
  Bidin}  \& {Mu{\~n}oz}}{{Villanova} et~al.}{2013}]{Villanova2013}
{Villanova} S.,  {Geisler} D.,  {Carraro} G.,  {Moni Bidin} C.,   {Mu{\~n}oz}
  C.,  2013, \mn@doi [\apj] {10.1088/0004-637X/778/2/186}, \href
  {https://ui.adsabs.harvard.edu/abs/2013ApJ...778..186V} {778, 186}

\bibitem[\protect\citeauthoryear{{Wardlow} et~al.,}{{Wardlow}
  et~al.}{2017}]{wardlow2017}
{Wardlow} J.~L.,  et~al., 2017, \mn@doi [\apj] {10.3847/1538-4357/837/1/12},
  \href {https://ui.adsabs.harvard.edu/abs/2017ApJ...837...12W} {837, 12}

\bibitem[\protect\citeauthoryear{{Weaver}, {McCray}, {Castor}, {Shapiro}  \&
  {Moore}}{{Weaver} et~al.}{1977}]{Weaver1977}
{Weaver} R.,  {McCray} R.,  {Castor} J.,  {Shapiro} P.,   {Moore} R.,  1977,
  \mn@doi [\apj] {10.1086/155692}, \href
  {https://ui.adsabs.harvard.edu/abs/1977ApJ...218..377W} {218, 377}

\bibitem[\protect\citeauthoryear{{Whitmore}, {Zhang}, {Leitherer}, {Fall},
  {Schweizer}  \& {Miller}}{{Whitmore} et~al.}{1999}]{Whitmore1999AJ}
{Whitmore} B.~C.,  {Zhang} Q.,  {Leitherer} C.,  {Fall} S.~M.,  {Schweizer} F.,
    {Miller} B.~W.,  1999, \mn@doi [\aj] {10.1086/301041}, \href
  {https://ui.adsabs.harvard.edu/abs/1999AJ....118.1551W} {118, 1551}

\bibitem[\protect\citeauthoryear{{W{\"u}nsch}, {Palou{\v{s}}}, {Tenorio-Tagle}
  \& {Ehlerov{\'a}}}{{W{\"u}nsch} et~al.}{2017}]{Wunsch2017}
{W{\"u}nsch} R.,  {Palou{\v{s}}} J.,  {Tenorio-Tagle} G.,   {Ehlerov{\'a}} S.,
  2017, \mn@doi [\apj] {10.3847/1538-4357/835/1/60}, \href
  {https://ui.adsabs.harvard.edu/abs/2017ApJ...835...60W} {835, 60}

\bibitem[\protect\citeauthoryear{{Yaghoobi} \& {Shadmehri}}{{Yaghoobi} \&
  {Shadmehri}}{2018}]{Yaghoobi2018MNRAS}
{Yaghoobi} A.,  {Shadmehri} M.,  2018, \mn@doi [\mnras] {10.1093/mnras/sty623},
  \href {https://ui.adsabs.harvard.edu/abs/2018MNRAS.477..412Y} {477, 412}

\bibitem[\protect\citeauthoryear{{Yaghoobi}, {Calura}, {Rosdahl}  \&
  {Haghi}}{{Yaghoobi} et~al.}{2022}]{Yaghoobi2022}
{Yaghoobi} A.,  {Calura} F.,  {Rosdahl} J.,   {Haghi} H.,  2022, \mn@doi
  [\mnras] {10.1093/mnras/stab3682}, \href
  {https://ui.adsabs.harvard.edu/abs/2022MNRAS.510.4330Y} {510, 4330}

\bibitem[\protect\citeauthoryear{{Zhu}, {Seaquist}  \& {Kuno}}{{Zhu}
  et~al.}{2003}]{zhu2003}
{Zhu} M.,  {Seaquist} E.~R.,   {Kuno} N.,  2003, \mn@doi [\apj]
  {10.1086/368353}, \href
  {https://ui.adsabs.harvard.edu/abs/2003ApJ...588..243Z} {588, 243}

\bibitem[\protect\citeauthoryear{{de Mink}, {Pols}, {Langer}  \& {Izzard}}{{de
  Mink} et~al.}{2009}]{Mink2009}
{de Mink} S.~E.,  {Pols} O.~R.,  {Langer} N.,   {Izzard} R.~G.,  2009, \mn@doi
  [\aap] {10.1051/0004-6361/200913205}, \href
  {https://ui.adsabs.harvard.edu/abs/2009A&A...507L...1D} {507, L1}

\makeatother
\end{thebibliography}

\appendix

\section{\INVH{}}\label{sec:Appen}
Here we test the formation of SG stars in a very dense medium with a density of $10^{-22} {\rm gcm^{-3}}$, which is similar to the density of a merging system such as the Antennae \citep{zhu2003}. The stalled radius of the bubble formed by the FG SN explosions for this simulation is found to be about $170\pc$ (from Equation $2$ in \citetalias{calura19}), comparable to the box size, applying a shorter $\tinf$ (from the birth of the cluster) that we assume to be $39\Myr$. Therefore we design the simulation such that the infall reaches the centre at about $39\Myr$ and therefore both the infall and the AGB ejecta are present at the start of the simulation.  

\begin{figure}
\centering
\includegraphics[width=\linewidth]{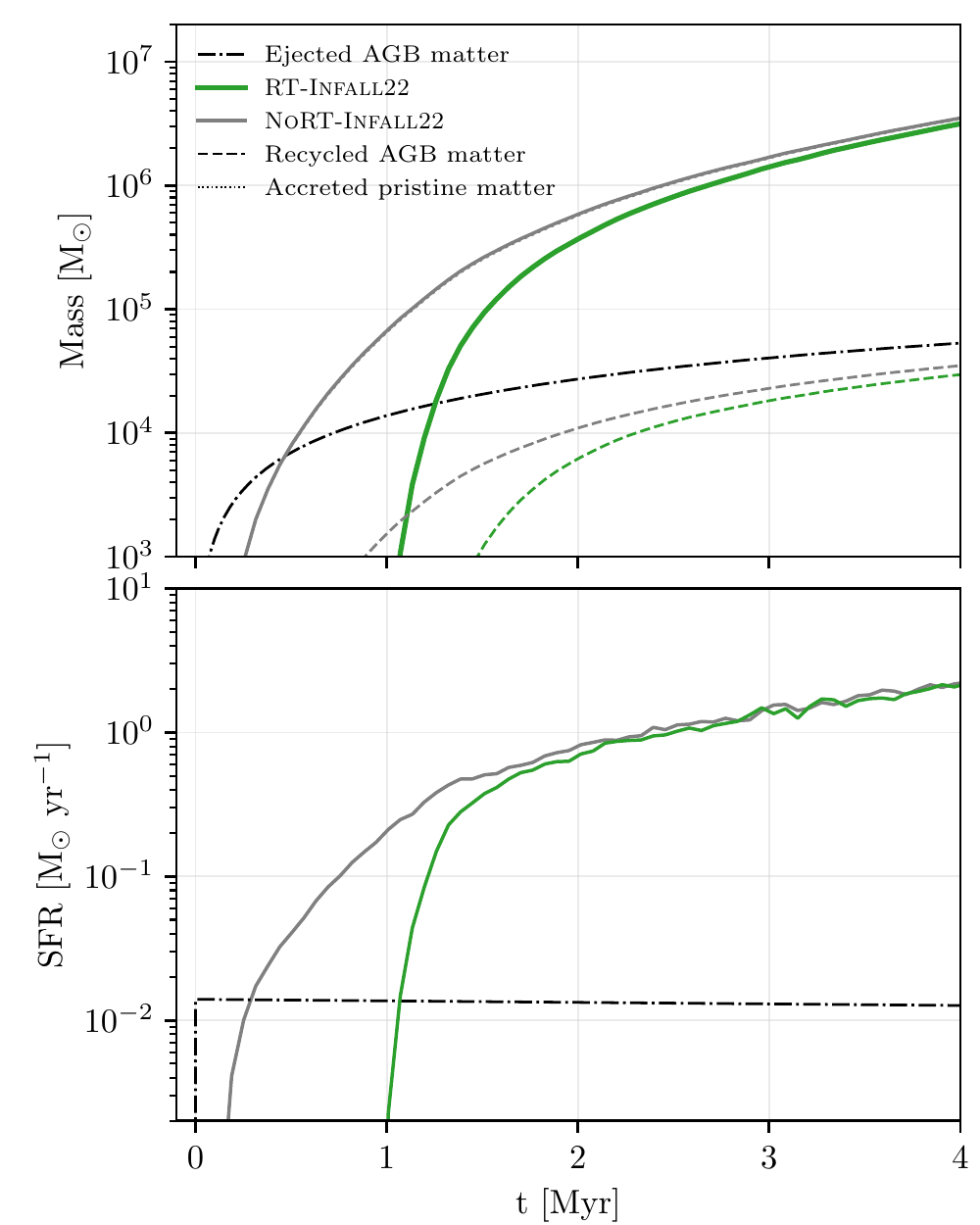}
\caption{As Fig. \ref{fig:SFR} but for the \INVH{} simulation.}
\label{fig:SFR22}
\end{figure}

\figref{fig:SFR22} shows the cumulative SG mass and the star formation rate versus time within the first $4\Myr$ for the \RT and \NONRT runs. The simulation experiences a SFR of about $2\Msun yr^{-1}$ after $4\Myr$. At such a high SFR it is expected that the number of massive stars considerably increases with respect to a standard IMF \citep[][also see \secref{SF}]{Tereza2018} and then after a few $\Myrs$ the SNe incorporated with SG stars will suppress the star formation. Hence, it may not be realistic that the SF goes on for a longer time. That is the reason why we stopped both simulations at $4\Myr$. However, it is not clear yet what is the efficiency of SNe in such a high-density medium and therefore a more comprehensive setup is required to investigate that. Nevertheless, we can still examine the effect of radiation in the ignition of star formation at early times.

Top panel of \figref{fig:SFR22} shows that the stars immediately start forming in the \NONRT simulation as the infall passes through the cluster, while star formation occurs $1\Myr$ later in the case of the \RT simulation. The difference between these two simulations disappears $\approx3\Myr$ after our time reference and finally the stellar mass ends up being about $3\times10^6 \Msun$ at $t=4\Myr$. 

But the SFRs in the \NONRT and \RT simulations converge at $ \approx2\Myr$. Here the  SFR has become $\sim 10$ times higher than the rate at the same time in the \INH as expected. 

\begin{figure}
\centering
\includegraphics[width=\linewidth]{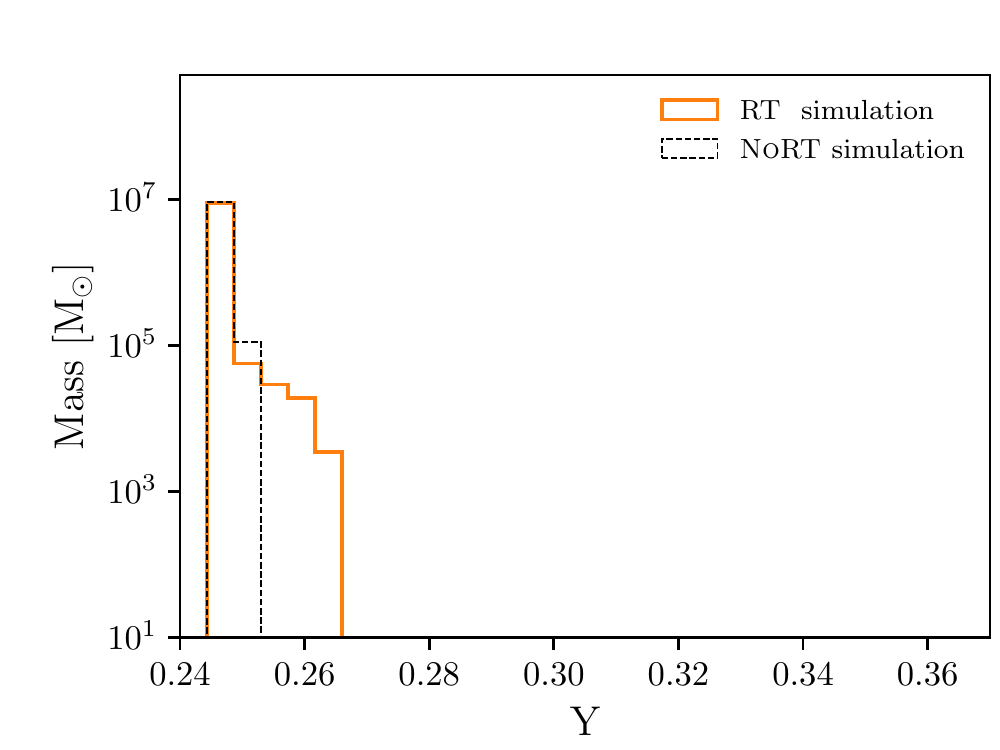}
\caption{ As Fig. \ref{fig:profs} but for the \INVH{} simulation.}
\label{fig:Hedist22}
\end{figure}

\figref{fig:Hedist22} shows that the maximum He abundance at $4\Myr$ is found to be about $0.27$, which is the lowest value compared to the other simulations due to very strong dilution of the AGB ejecta in this run. If we allow the simulation to proceed for a longer time,  this maximum value does not change since new stars would have the same He abundance as in the pristine gas, resulting from a high accretion and dilution. Therefore the results of this model present a He enrichment of about $0.02$, which is much lower than the observed values for very massive clusters \citep{milone2020}.

\bsp	
\label{lastpage}
\end{document}